\documentclass[a4paper,twocolumn,11pt]{quantumarticle}
\pdfoutput=1
\usepackage[utf8]{inputenc}
\usepackage[english]{babel}
\usepackage[T1]{fontenc}
\usepackage{amsmath}
\usepackage{hyperref}
\usepackage{amsfonts,breqn,graphicx,braket,bm}

\usepackage{tikz}
\usepackage{lipsum}

\usepackage{algorithm}
\usepackage{algpseudocode}

\begin{document}

\title{Multiobjective variational quantum optimization for constrained problems: an application to Cash Management}

\author{Pablo Díez-Valle}
\email{pablo.diez@csic.es}
\orcid{0000-0001-8338-7973}
\affiliation{Instituto de F\'{\i}sica Fundamental IFF-CSIC, Calle Serrano 113b, Madrid 28006, Spain}
\author{Jorge Luis-Hita}
\affiliation{BBVA Quantum, Calle Azul 4, 28050 Madrid, Spain}
\author{Senaida Hernández-Santana}
\affiliation{BBVA Quantum, Calle Azul 4, 28050 Madrid, Spain}
\author{Fernando Martínez-García}
\affiliation{Instituto de F\'{\i}sica Fundamental IFF-CSIC, Calle Serrano 113b, Madrid 28006, Spain}
\author{\\Álvaro Díaz-Fernández}
\affiliation{BBVA Quantum, Calle Azul 4, 28050 Madrid, Spain}
\author{Eva Andrés}
\affiliation{BBVA Quantum, Calle Azul 4, 28050 Madrid, Spain}
\author{Juan José García-Ripoll}
\affiliation{Instituto de F\'{\i}sica Fundamental IFF-CSIC, Calle Serrano 113b, Madrid 28006, Spain}
\author{Escolástico Sánchez-Martínez}
\affiliation{BBVA Quantum, Calle Azul 4, 28050 Madrid, Spain}
\author{Diego Porras}
\affiliation{Instituto de F\'{\i}sica Fundamental IFF-CSIC, Calle Serrano 113b, Madrid 28006, Spain}

\maketitle

\begin{abstract}
Combinatorial optimization problems are ubiquitous in industry. In addition to finding a solution with minimum cost, problems of high relevance involve a number of constraints that the solution must satisfy. Variational quantum algorithms have emerged as promising candidates for solving these problems in the noisy intermediate-scale quantum stage. However, the constraints are often complex enough to make their efficient mapping to quantum hardware difficult or even infeasible. An alternative standard approach is to transform the optimization problem to include these constraints as penalty terms, but this method involves additional hyperparameters and does not ensure that the constraints are satisfied due to the existence of local minima. In this paper, we introduce a new method for solving combinatorial optimization problems with challenging constraints using variational quantum algorithms. We propose the Multi-Objective Variational Constrained Optimizer (MOVCO) to classically update the variational parameters by a multiobjective optimization performed by a genetic algorithm. This optimization allows the algorithm to progressively sample only states within the in-constraints space, while optimizing the energy of these states. We test our proposal on a real-world problem with great relevance in finance: the Cash Management problem. We introduce a novel mathematical formulation for this problem, and compare the performance of MOVCO versus a penalty based optimization. Our empirical results show a significant improvement in terms of the cost of the achieved solutions, but especially in the avoidance of local minima that do not satisfy any of the mandatory constraints.  
\end{abstract}

\section{Introduction}

Quantum computing holds the promise of a major impact on science and industry due to its capacity to solve complex problems. Some potential applications are the quantum simulation of many-body problems~\cite{Reiher_2017,von_Burg_2021}, numerical analysis tasks such as solving differential equations~\cite{Lubasch_2020,Garc_a_Molina_2022}, machine learning~\cite{Biamonte_2017,Benedetti_2019,Perdomo-Ortiz_2018}, etc. A widely studied area among them is combinatorial optimization (CO) which consists of searching for an optimal solution among a finite set of elements~\cite{Nikolaj2018}. Finding the exact solution to these problems is in many cases a well-known NP-hard problem~\cite{Nemhauser1988IntegerAC}, i.e., it is not possible in polynomial time using classical computing. Some canonical examples are the traveling salesman problem \cite{Lenstra1975} or max-cut \cite{Goemans1995ImprovedAA,Festa2002}. Although quantum computers may not be able to solve this task exactly in polynomial time either, heuristic quantum algorithms may still be able to achieve better approximate solutions to these problems in terms of speed, quality of the obtained solutions, or even resource savings. One of the leading quantum paradigms to tackle these problems in the near term are the Variational Quantum Algorithms (VQA) \cite{Cerezo_2021}. Such algorithms use a hybrid quantum-classical approach to approximate the minimum energy state of a Hamiltonian that encodes our problem. The combination of quantum and classical computing allows the use of shallow quantum circuits suitable for the noisy intermediate-scale quantum (NISQ) era \cite{Preskill_2018,Bharti_2022}.  

CO arises in a wide range of industrial domains such as finance~\cite{Dominguez2017}, logistics~\cite{Sbihi2010CombinatorialOA}, artificial intelligence~\cite{Bezerra2021}, supply chain~\cite{ESKANDARPOUR201511} or drugs discovery~\cite{Kennedy2008ApplicationOC}. Therefore, even if they are potentially only approximations of the global optimum, better CO solutions have a significant practical value. Real-world combinatorial optimization problems usually involve not only the minimization of a cost function, but also a number of equality and inequalities hard constraints that must be satisfied by feasible solutions. In the context of quantum optimization, a desirable technique would be to implement constrained algorithms such as the Quantum Alternating Operator Ansatz, with quantum circuits capable of natively preserve the constraints~\cite{Hadfield2017QuantumAO,Hadfield_2019}. Some work has been done in this direction, including recent experimental demonstrations on text summarization \cite{niroula2022constrained}. However, the constraints of realistic formulations of practical problems are often too difficult to be efficiently mapped to a quantum processor. In this scenario, the state-of-the-art strategy to force constraint satisfaction is to transform the original cost function by introducing penalty terms that artificially raise the energy of infeasible solutions \cite{lucas2014ising}. Although easily implementable, this technique does not guarantee the convergence of the algorithm to a solution that satisfies all constraints, and suffers from shortcomings such as the adjustment of instance-dependent penalty hyperparameters. Despite their critical relevance in practical scenarios, few studies have been conducted to explore new possibilities for general constraint encoding in VQAs. Indeed, studies on the performance of VQAs are usually performed using well-studied theoretical problems that, although very valuable in addressing the potential of quantum algorithms, lack direct practical applicability in the industrial sector.    

In this work, we propose the multiobjective variational constrained optimizer (MOVCO), a method for improving the convergence of variational quantum algorithms to optimal solutions satisfying a set of restrictions, and test its performance on an industrially relevant problem. MOVCO relies on a genetic algorithm to simultaneously optimize the projection of the variational wavefunction onto the subspace of solutions satisfying all constraints, and the energy of the feasible solutions. This is, to the best of our knowledge, one of the first papers addressing a variational method for the efficient optimization of realistic problems involving a large number of general constraints. Very recently, during the preparation of this manuscript, a study has been released that introduces a modification to the objective function of VQAs to deal with hard constraints~\cite{Hao2022}. Specifically, they propose the minimization of an in-constraint energy instead of the penalized cost function. However, despite also being one of the only papers discussing this question, our multiobjective approach is clearly different.

The manuscript is structured as follows. First, we provide a brief review of key concepts to understand the new method, such as Variational Quantum Algorithms (sec.~\ref{sec_VQAs}), constrained Combinatorial Optimization (sec.~\ref{sec_CO}), and the Non-dominated Sorting Algorithm (sec.~\ref{sec_NGSAII}). We introduce MOVCO in section~\ref{sec_MOVCO}, and pose the strengths of the method. We then demonstrate the effectiveness of the method on a real-world financial and logistical problem known as Cash Management problem. We describe in detail the specifications of this practical problem in section \ref{sec_CMdescription}, and propose a novel mathematical formulation in terms of spins in section \ref{sec_CMformulation}. In section \ref{sec_numresults} we show the empirical results that support an advantage of MOVCO to avoid infeasible local minima. We conclude with some remarks on the importance of the work, future lines of research, and potential improvements of MOVCO. 

\section{Background}
%
%
%

\subsection{Variational quantum optimization}
\label{sec_VQAs}
%
%
%
%

Variational quantum algorithms (VQA) are a class of hybrid quantum-classical algorithms widely studied due to their ability to solve complex problems with shallow quantum circuits \cite{Cerezo_2021}. Their potential is based on the combination of powerful classical optimization methods with the suitability of quantum circuits to explore the entire Hilbert space efficiently. There is a great variety of VQAs whose features are adapted to the intended application. Nevertheless, they have some common ingredients and operate according to the following scheme. 

VQAs rely on a quantum computer to build an ansatz through the application of a sequence of quantum operations $U(\vec{\theta})$ to the input states: 
\begin{equation}
 \ket{\Psi (\vec{\theta})} = U(\vec{\theta}) \ket{0}^{\otimes N},
\label{eq_ansatz}
\end{equation}
where $N$ is the number of qubits. These operations depend on a number of tunable parameters $\vec{\theta}$. The unitary operations $U(\vec{\theta})$ can be expressed as a product of $L$ unitaries, where $L$ is the number of layers of the ansatz that determines the depth of the quantum circuit:
\begin{equation}
 U(\vec{\theta}) = \prod_{l=1}^{L} U_l(\vec{\theta}_l).
\label{eq_ansatz_layers}
\end{equation}
Then, the parameters of the ansatz are tuned according to a cost function that encodes the problem. This tuning is performed by a classical computer that takes measurements on the ansatz as input to find the optimal parameters that bring $\ket{\Psi (\vec{\theta})}$ as close as possible to the best solution.    

This kind of algorithms have been extensively studied in the context of combinatorial optimization~\cite{Nikolaj2018,Nannicini2019,Willsch_2020,Robert_2021,Amaro_2022,Azad_2022,Leontica2022}. Although there is still no proof that these algorithms can provide a quantum advantage in optimization problems, there are indications of their potential~\cite{Crooks2018,Streif2020,Diez2022,Farhi_2022}. Examples of VQAs that have been studied in this context are the Quantum Approximate Optimization Algorithm (QAOA) \cite{Farhi_2014}, Variational Quantum Eigensolver (VQE) \cite{Peruzzo_2014}, recursive QAOA (RQAOA) \cite{Bravyi2020}, Filtering VQE (F-VQE) \cite{Amaro_2022_FVQE}, Layer VQE (L-VQE) \cite{Liu_2022}, ADAPT-QAOA \cite{Zhu_2020}, Variational Quantum Imaginary Time Evolution (VarQITE) \cite{McArdle_2019}, etc. In this manuscript, we will focus our numerical experiments on the Variational Quantum Eigensolver, whose problem-agnostic construction can be well adapted to a black-box optimization such as the Cash Management problem introduced in section \ref{sec_cashmanagement}. However, the new multiobjective technique presented in this paper can be easily extended to any of these variational algorithms. 

\subsubsection{Variational Quantum Eigensolver}
\label{sec_vqe}

\begin{figure}
  \centering
\includegraphics[width=\linewidth]{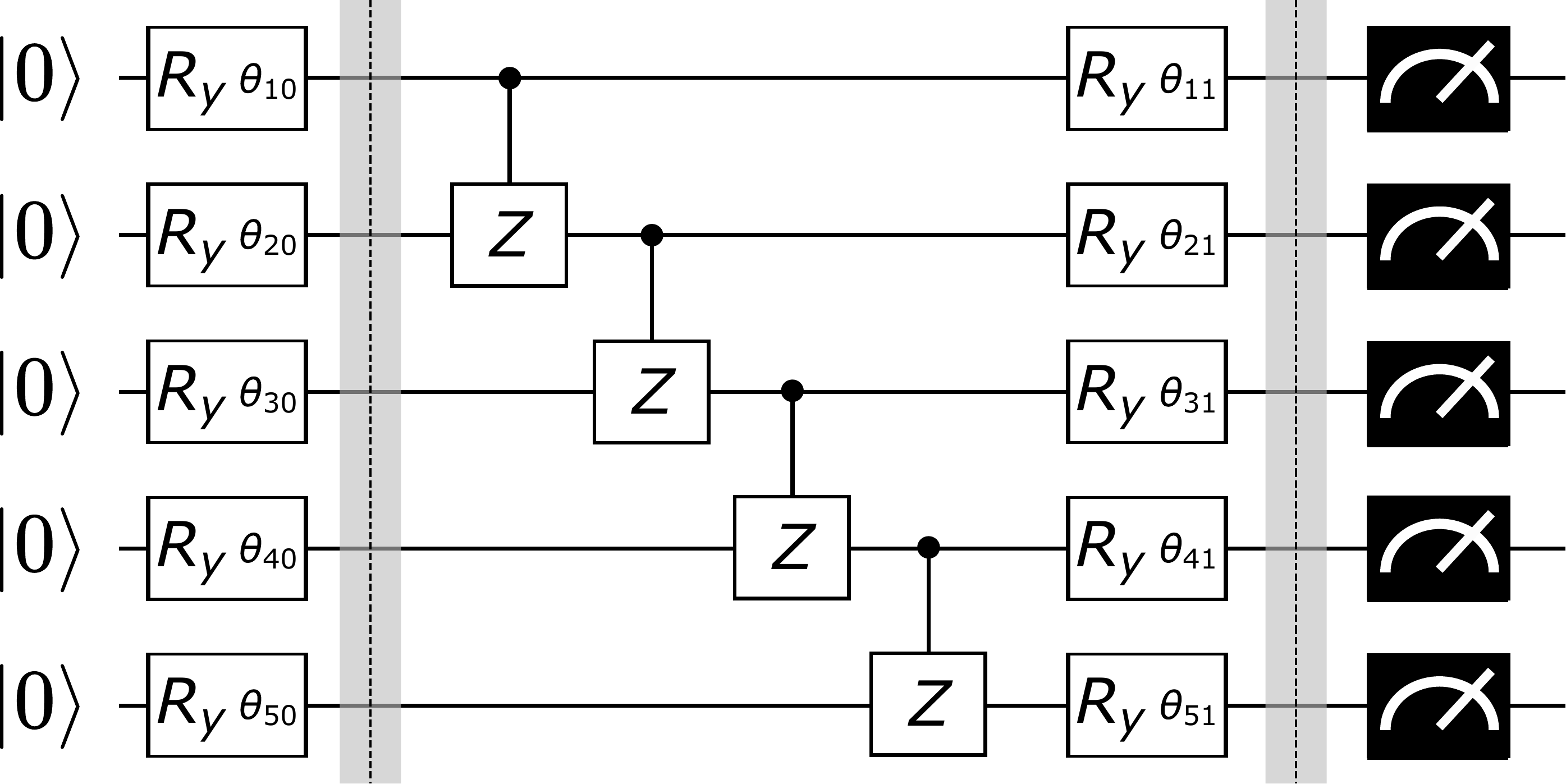}
\caption{One layer of the harware-efficient variational quantum circuit~\eqref{eq_ryansatz} for five qubits. We apply parameterized single qubit $R_y$ rotations and control-Z entangling gates between first-neighbors.}
  \label{fig_ryansatz}
\end{figure}

The VQE was originally proposed as a variational algorithm for finding the ground state energy of a chemical molecule described by a Hamiltonian $\hat{H}$ \cite{Peruzzo_2014}, but it has also been applied in the context of quantum optimization \cite{Amaro_2022}. 

In VQE, the parameters of an ansatz are trained by the minimization of the expectation value of the Hamiltonian $\bra{\Psi (\vec{\theta})}\hat{H}\ket{\Psi (\vec{\theta})}$. Although some problem-dependent or problem-inspired ansätze for VQE has been proposed to solve problems in quantum chemistry~\cite{Lee_2018,Wecker_2015,Wiersema_2020}, in this algorithm the variational ansatz~\eqref{eq_ansatz} is free, in the sense that it does not need to be determined by the Hamiltonian of the problem as in QAOA. In particular, a typical hardware-efficient choice is the sequence of quantum gates shown in Fig.\ref{fig_ryansatz}:
\begin{equation}
U(\vec{\theta}) =
\prod_{l=1}^{L}\left[\prod_{n=1}^{N}R_y (\theta_{nl}) U_{\textnormal{ent}}\right]\prod_{n=1}^{N} R_y (\theta_{n0})\,, 
\label{eq_ryansatz}
\end{equation}
where $R_y (\theta_{nl}) \equiv e^{i\theta_{nl}\hat{Y}_n}$ are single-qubits rotations around the y axis, $U_{\textnormal{ent}} \equiv \prod_{n=1}^{N-1} e^{i\frac{\pi}{4}\left(\mathbb{I}-\hat{Z}_n\right)\left(\mathbb{I}-\hat{Z}_{n+1}\right)}$ is an entangling gate made up of two-qubit control-Z gates between each qubit with its nearest-neighbor qubit in a linear quantum processor topology, and $\hat{Y}_n$ and $\hat{Z}_n$ are the Pauli Y and Z operators acting on the n-th qubit. In the computational experiments shown in section~\ref{sec_numresults}, the parameters $\theta_{n0}$ are initialized around $\sim\pi/4$ so that the initial state is close to the full superposition, with small random perturbations $\theta_{nl}\sim10^{-2}$.  

The flexibility and first-neighbor connection of hardware-efficient ansätze make them suitable for the realization of this algorithm on real quantum processors. However, they are not tailored to the problem and they may suffer from trainability issues such as barren plateaus~\cite{McClean_2018}. 

\subsection{Constrained combinatorial optimization}
\label{sec_CO}

Combinatorial optimization problems with relevance in industry usually involve the satisfaction of a large number of hard constraints. These problems can be formulated as the minimization of a black-box cost function,
\begin{equation}
    \min C(z) \,\,\, \textnormal{with}\,\, z\in\{-1,+1\}^N \,,
\end{equation}
such that the solution $z$ must fulfill a number of inequality and equality constraints:
\begin{equation}
   b_i(z) = 0 \,\,,\,\,g_j(z) \leq 0 
   \label{eq_classicalconstraints}
\end{equation}
where $N$ is the number of variables of the problem, and $i = 1,...,I$ , $j = 1,...,J$, with $I \in \mathbb{N}$ and $J \in \mathbb{N}$ the number of equality and inequality constraints, respectively. This cost function can be translated to a Hamiltonian $\hat{C}$ through the change of variables $z_n\rightarrow \hat{Z}_n$, where $\hat{Z}_n$
is the Pauli $Z$ matrix acting on the $n$-th qubit. Thereby, the problem becomes the search of the ground-state of this Hamiltonian. Regarding the variational algorithms, this is equivalent to finding the parameters $\vec{\theta}$ that bring the wavefunction closer to that ground state:   
\begin{equation}
    \underset{\vec{\theta}}{\textnormal{min}} \left[\bra{\Psi(\vec\theta)}\hat{C}(\hat{Z})\ket{\Psi(\vec\theta)}\right] \,,
\end{equation}
such that
\begin{equation}
\begin{split}
\bra{\Psi(\vec\theta)}\hat{B}_i(\hat{Z})\ket{\Psi(\vec\theta)} = 0 \;,\\ \bra{\Psi(\vec\theta)}\hat{G}_j(\hat{Z})\ket{\Psi(\vec\theta)} \leq 0 \,, 
   \label{eq_constraints}
\end{split}
\end{equation}
where the constraints~\eqref{eq_classicalconstraints} are also formulated in terms of operators $\hat{B}_i$, $\hat{G}_j$ through the previous transformation.
We denote the subspace composed of the states satisfying~(\ref{eq_constraints}) as the \textit{in-constraint} or \textit{feasible subpace} $\mathcal{S}$. In a realistic scenario, the trial state $\ket{\Psi(\vec\theta)}$ is prepared in an $N$ qubits quantum computer so that we do not have access to the exact probability amplitudes of the wavefunction. The average energy $\bra{\Psi(\vec\theta)}\hat{C}(\hat{Z})\ket{\Psi(\vec\theta)}$ is then estimated simultaneously measuring $K$ times all qubits in the Pauli $z$ basis. Each measure produces a classical state $\ket{\Psi_k} = \{-1,+1\}^N$ whose cost function value $C_k = \bra{\Psi_k}\hat{C}\ket{\Psi_k}$ is efficiently computed. Hence, the sample mean is the estimator that we actually minimize,
\begin{equation}
    \bra{\Psi(\vec\theta)}\hat{C}(\hat{Z})\ket{\Psi(\vec\theta)}\approx\underset{\vec{\theta}}{\textnormal{min}} \left[\dfrac{1}{K}\sum_{k=1}^K \hat{C}_k(\vec\theta)\right]\,,
    \label{eq_costestimator}
\end{equation}
where $k=1,...,K$ denotes each of the $K$ samples measured. Therefore, constrained variational quantum optimization aims to approximate a wavefunction able to sample classical states with low $C_k$ such that $\bra{\Psi_k}\hat{B}_i(\hat{Z})\ket{\Psi_k} = 0$ and $\bra{\Psi_k}\hat{G}_j(\hat{Z})\ket{\Psi_k} \leq 0$.    

\subsection{Non-dominated Sorting Genetic Algorithm}
\label{sec_NGSAII}

\begin{figure*}
  \centering
\includegraphics[width=1\linewidth]{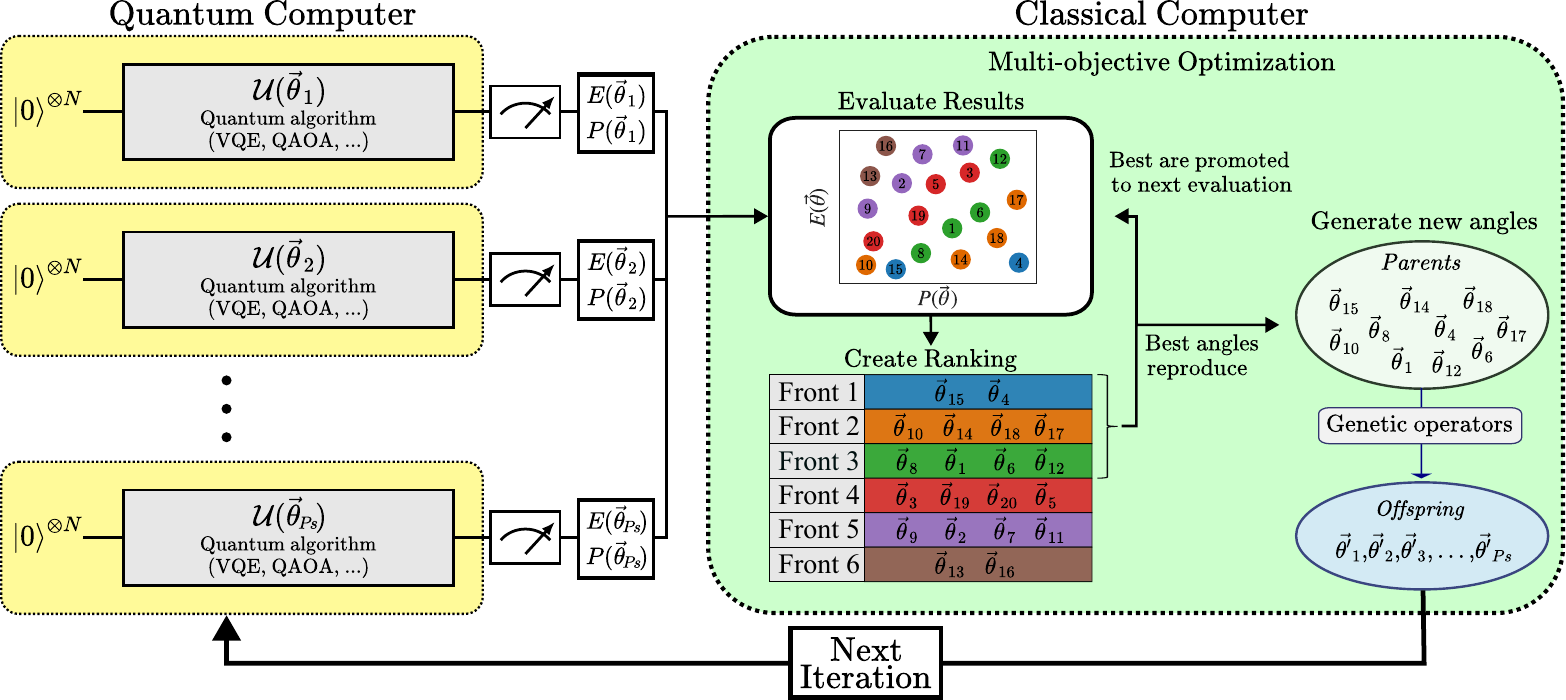}
\caption{Schematic diagram of the constrained multi-objective variational optimizer (MOVCO). The hybrid algorithm is composed of a variational ansatz $\mathcal{U}(\vec{\theta})$ running on a quantum computer, and a multiobjective optimization performed by the classical genetic method NSGA-II. Initially, the quantum circuit $\mathcal{U}$ is sampled with a collection of sets of random angles $\vec{\theta}_i$ to compute the fitness functions \eqref{eq_fitnessprojection} and \eqref{eq_restrictedenergy}, that address the quality of the solutions in terms of constraint satisfaction and energy optimization. These results are ranked in terms of Pareto fronts so that angles belonging to the first fronts are saved for the next iteration (parent population). A collection of new sets of angles (offspring population) is generated from the application of genetic operators (tournament selection, crossover, and mutation) on the parent population. In the next iteration, the quantum circuit $\mathcal{U}$ is sampled with only the offspring parameters, performing the multi-objective optimization with the parent angles from the previous iteration and the new offspring results (Color online).}
  \label{fig_MOVCOscheme}
\end{figure*}

The Non-dominated Sorting Genetic Algorithm (NSGA-II), introduced in \cite{ngsaII}, is an evolutionary algorithm to perform multiobjective optimization which has been proven to be suitable for large-scale optimization problems \cite{Deb2011,nebro2022}. The ultimate goal of NSGA-II is to find a set of optimal solutions for multiple cost functions. This set is known as the set of non-dominated solutions, or \textit{Pareto Front}. A solution $x_0$ is said to be non-dominated if and only if there is no different solution $x_i\neq x_0$ such that $x_i$ is better than $x_0$ in at least one objective function, and at the same time is equal to or better than $x_0$ for all other cost functions being optimized.

NSGA-II belongs to the subset of evolutionary algorithms known as genetic algorithms, which are meta-heuristic optimization techniques based on biological-inspired operators such as natural selection, crossover, and mutations \cite{Katoch_2021}. The optimization process is carried out by iteratively evolving a population (a set of solutions) to obtain better individuals (better solutions) using the concept of survival of the fittest and the previous operators. Specifically, NSGA-II involves the following steps (see Appendix~\ref{append_NGSAIIpseudocodes} for more details):
\begin{enumerate}
    \item \textit{Population initialization}: the algorithm is initialized with a set of randomly sampled solutions (individuals). After the first iteration, the population will be a combination of the parent and offspring populations from the previous iteration. The population size is a hyperparameter of the model.
    
    \item \textit{Non-dominated sorting}: the individuals of the population are sorted and classified in \textit{fronts} according to its Pareto dominance. In other words, individuals which are non-dominated by any other element are assigned to front 1 and eliminated from consideration. Then, the non-dominated individuals from the remaining population are classified as front 2, and so on with all the elements of the population.
    
    \item \textit{Create parents population}: the new population is created from the front classification so that individuals belonging to the first fronts are promoted to the next iteration.
    
    \item \textit{Crowding distance sorting}: when a front is partially taken, i.e. not all elements of a front are needed to fill the parent population, the solutions of this front are sorted according to a density-based metric. To encourage a broader exploration of the solution space, individuals in less dense spaces are promoted. 
    
    \item \textit{Create offspring population}: using genetic operators a new population is generated from the parent population. First, a binary tournament selection is performed so that only the best individuals (based on front and crowding distance ranking) from a random sampling are allowed to reproduce. Then, the variables (genes) of two of these individuals (parents) are combined following a certain crossover rule and giving rise to a new solution (child). Lastly, a mutation may be applied to the new individuals, changing some of their variables according to a chosen probability.
    
    \item \textit{Iterate the process}: steps (ii) to (iv) are repeated with a new population composed of the parents and offspring populations. Because of the genetic simile, each iteration of the algorithm is called a generation.
    
\end{enumerate}

In our numerical experiments we use the NGSA-II implementation of the open-source Python framework \textit{pymoo 0.5.0} \cite{pymoo}.

\section{Variational constraint optimization with multiobjective cost}
\label{sec_MOVCO}

\begin{figure}
  \centering
\includegraphics[width=\linewidth]{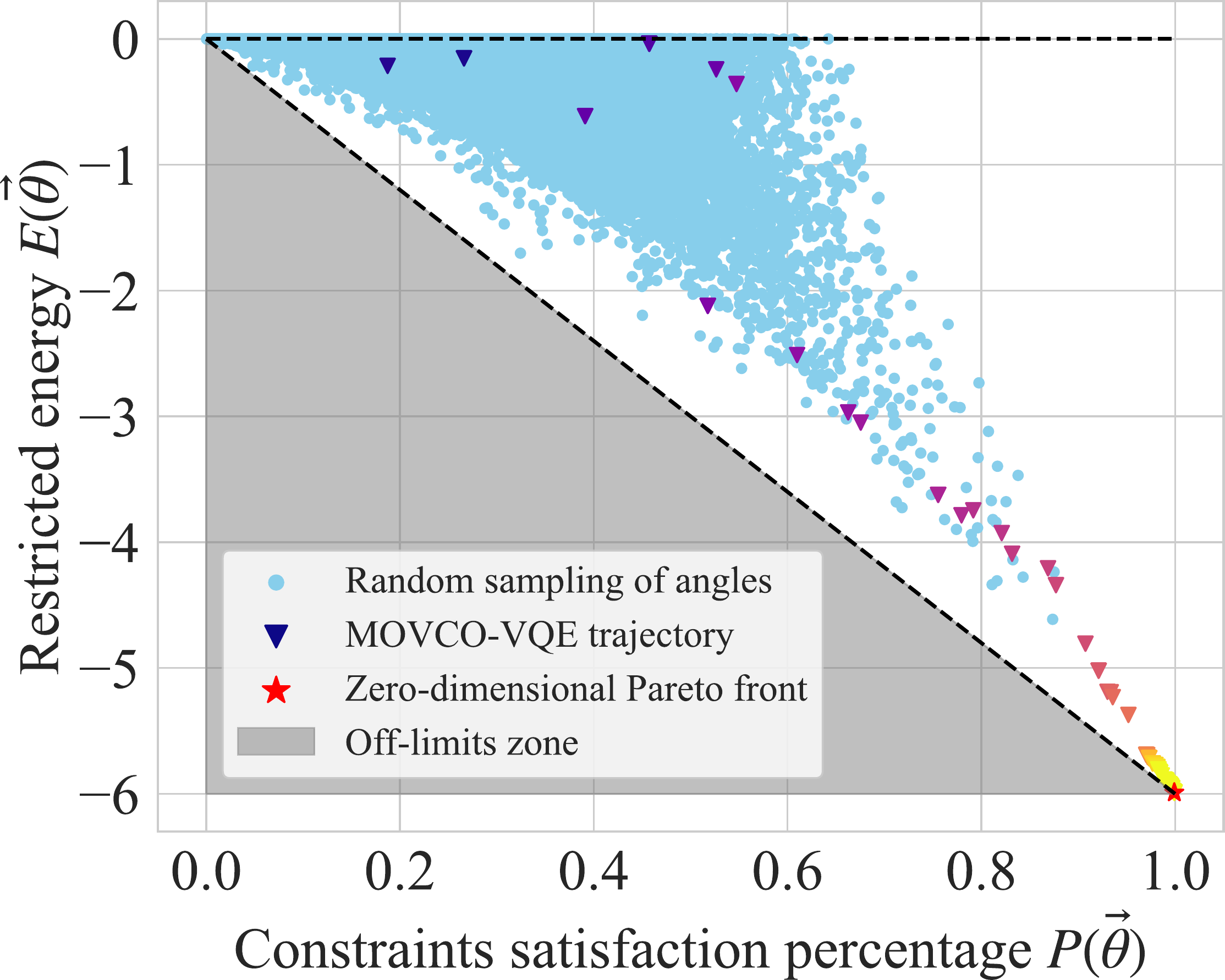}
\caption{Optimization landscape in the space of fitness functions for an 8-qubits CMP instance with 2 cash points optimized in a 2 days period as explained in section \ref{sec_cashmanagement}. The maximum cost of transactions is $\sum_c k^0_c + \sum_c k_c = 18$, and the minimum cost that satisfies all constraints 12. Therefore the zero-dimensional Pareto front is $\left(P,E\right)_{PF}=\left(1,-6\right)$. We trace the trajectory of optimal solutions found in each generation of MOVCO-VQE. We also show the results with $10^5$ set of variational angles drawn from a random uniform sampling, and highlight the region where the variational wavefunction cannot sample (Color online).}
  \label{fig_MOVQElandscape}
\end{figure}

We now introduce the \textit{multiobjective variational constrained optimizer} (MOVCO), a new method to solve combinatorial optimization problems with hard constraints which combines the quantum variational framework with a genetic multiobjective optimization such as NSGA-II, where each individual is a set of variational parameters $\vec{\theta}$ (see Fig.\ref{fig_MOVCOscheme}). In this algorithm the parameters of a variational wavefunction $\ket{\Psi (\vec{\theta})}$ are iteratively updated through the simultaneous optimization of two fitness functions: one addresses the quality of the solutions in terms of the constraints satisfaction, and the other deals with the energy optimization, but preserving the variational algorithm from consuming time optimizing solutions outside the in-constraints regime.    

\begin{itemize}
    \item \textit{Fitness function to maximize the constraints satisfaction.-}
Ideally, we would be interested in directly maximizing the projection of the wavefunction $\ket{\Psi (\vec{\theta})}$ on the feasible space, i.e, the subspace of solutions that satisfy the constraints. However, this calculation is impractical in a realistic scenario since we do not even generally know which is the feasible space for large-size systems. All we can compute efficiently is whether or not a given solution satisfies given constraints, i.e, we can know if the solution belongs to such a subspace. Furthermore, if we only maximize solutions with full overlap on the feasible space, we would easily reach a stagnation since that subspace will presumably be small in comparison with the whole Hilbert space. Instead, we maximize the percentage of constraints that are satisfied by the variational ansatz.  

\begin{equation}
    P(\vec{\theta}) = \frac{1}{K} \sum_{k=1}^{K} P_k(\vec{\theta})\,\,,
\label{eq_fitnessprojection}
\end{equation}
where $P_k$ with $k = 1,...,K$ is the percentage of constraints satisfied by each sampled solution. 
    \item \textit{Fitness function to minimize the energy of feasible solutions.-}
Besides states that satisfy the constraints, the optimization problem aims to find the lowest energy solutions among them. To avoid wasting time sampling and optimizing from solutions that lay outside the feasible subspace, instead of the traditional average energy~(\ref{eq_costestimator}), we propose to simultaneously optimize the following \textit{restricted energy}:
\begin{equation}
    E(\vec{\theta}) = \sum_{k \in \mathcal{S}}\left(  C_k(\vec{\theta}) - \max\left[C\right] \right) / K
    \label{eq_restrictedenergy}
\end{equation}
where $C_k$ with $k = 1,...,K$ are the values of the costs calculated for each sampled solution, and $\mathcal{S}$ is the subspace of solutions that satisfy all constraints. Therefore, only the subset of the $K$ sampled states belonging to the feasible space $S$ is used to compute the restricted energy. $\max\left[C\right]$ is a non-strict upper bound to the cost function value that can be efficiently calculated in practical scenarios. For instance, given a QUBO problem $C = \sum_{i,j} z_i Q_{i j} z_j$ a straightforward upper-bound is $\max\left[C\right] = \sum_{i,j}|Q_{i j}|$. Note that this choice confines the restricted energy since 
\begin{equation}
    0 \geq E(\vec{\theta}) \geq \sum_{k \in \mathcal{S}} \frac{1}{K}  \left( \min\left[C\right] - \max\left[C\right] \right)\,,
\end{equation}
and $\min \left[E(\vec{\theta})\right] = \min \left[C\right]- \max \left[C\right]$, with $1\geq P(\vec{\theta})\geq\sum_{k \in \mathcal{S}} \frac{1}{K}$. The selection of specific suitable sampled states for computing the energy estimator of variational quantum algorithms is connected with other ideas in the literature such as the Conditional Value-at-Risk VQE (CVaR-VQE)~\cite{Barkoutsos_2020}, where only the lowest energy states are used in the energy calculation.      
\end{itemize}

The combination of these fitness functions creates a favorable landscape (see Fig.\ref{fig_MOVQElandscape}) that enhances the convergence of the algorithm to low-energy feasible solutions, as shown in section~\ref{sec_numresults}. The Pareto front that maximizes the projection on the feasible space and minimizes the restricted energy is zero-dimensional in the space of the functions: $\left(P,E\right)_{PF}=\left(1,\min E\right)$. This fact allows the algorithm to quickly increase $P$ in such a way that after a number of iterations the ansatz only samples feasible states, avoiding wasting time optimizing infeasible solutions. Furthermore, the optimized functions 
may be simultaneously measured in the quantum processor.   

One of the advantages of the multiobjective optimization implemented by MOVCO is the capability to globally explore the phase space, as opposed to the narrower scope of an optimization defined by one optimal trajectory. 
As explained in section~\ref{sec_NGSAII}, the genetic algorithm enables to sample states which may be far away in the variational parameter space, leading to a more efficient exploration of the Hilbert space. As an example, we can observe in Fig.\ref{fig_MOVQElandscape} how even a random sampling of angles allows reaching a wide area of the fitness functions space of the small size problem. This feature, together with the optimization of the projection~\eqref{eq_fitnessprojection}, provides a higher chance of convergence to the in-constraint subspace.  

 
We believe that this method is well suited for combinatorial optimization problems where a considerable number of solutions fulfill at least some of the constraints. Scenarios where the subspace of states satisfying none of the constraints is too large could result in effective barren plateaus where the restricted energy is trapped at 0, due to the difficulty of sampling a state within the feasible space $\mathcal{S}$. Nevertheless, many problems of industrial interest satisfy this requirement.

Note that this method targets the objective function used to optimize the variational parameters. Therefore, although the results shown in later sections are centered on VQE (MOVCO-VQE), our approach is ansatz agnostic and can be broadly applied to other variational algorithms such as QAOA or layer-VQE. Known modifications of objective functions, such as CVaR-VQE \cite{Barkoutsos_2020}, can also benefit from this method. In this case, the values of $C_k$ in~(\ref{eq_restrictedenergy}) would not be restricted to all states in $\mathcal{S}$, but only the lowest energy values among them would be taken, which may lead to faster convergence to solutions with a minimal overlap with the ground-state.  


\section{Application to a real-world problem: Cash Management}
\label{sec_cashmanagement}

In this section, we introduce the \textit{Cash Management problem} (CMP), a non-convex optimization problem of high industrial interest. We also introduce a novel mathematical formulation of this problem that scales linearly with the number of cash points and the length of the optimized temporal period.

\subsection{Description of the problem}
\label{sec_CMdescription}

\begin{figure}
  \centering
\includegraphics[width=\linewidth]{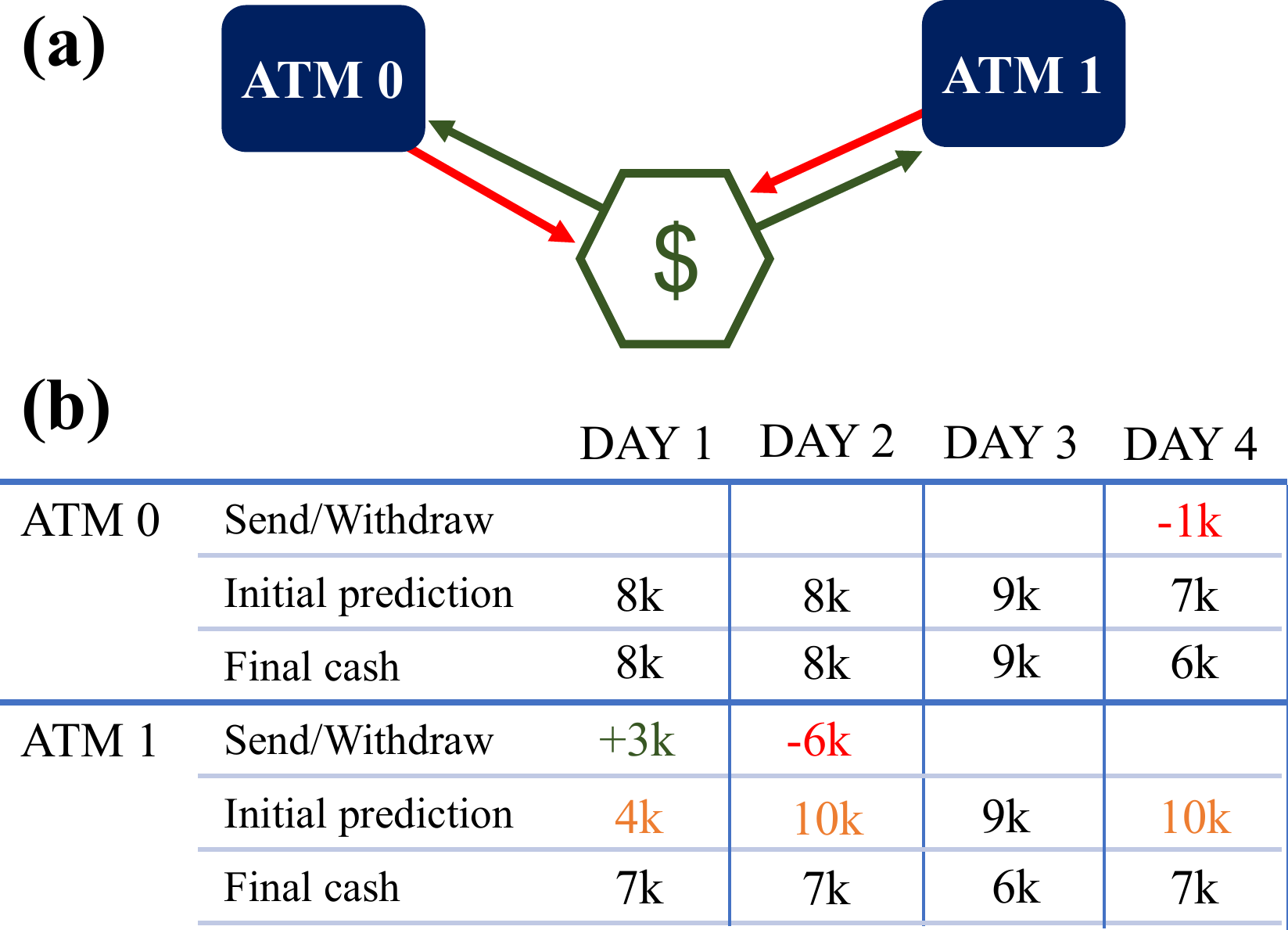}
\caption{(a) Scheme of the Cash Management problem. The money is supplied and withdrawn from a vault to ensure that an ATM network has the correct amount of cash. (b) Example of a CMP with two optimized ATMs over a four-day period. The cash in each ATM every day must be in the range $\left[\right.$6k,9k$\left.\right]$, and the sum of all money in the network must be less than 13k on the last day. Optimal cash transfers are scheduled to fulfill the above requirements (Color online).}
  \label{fig_CMPscheme}
\end{figure}

One of the tasks that are part of a bank's routine operations is planning cash deliveries to branches and automated teller machines (ATMs) to provide them with the cash they need on a daily basis. This planning is based on a forecast of the daily cash that will be available at each branch and ATM, and is subject to certain restrictions. For example, when the amount of cash at a given ATM is expected to be less than a certain pre-set minimum amount, a request is made to send cash to that branch. This money comes from the vaults in which the amounts of cash that may be needed in the next few days are stored (see Fig.~\ref{fig_CMPscheme}). 

The transport from the vault to the branch or ATM is usually carried out by an external company that has been requested to send or withdraw cash to a certain branch. The external company is then in charge of the transport, defining the route, which usually changes for security reasons. Thus, the Cash Management problem is not a generalization of the Traveling Salesman Problem as it might seem at first glance since we have no control over the specific delivery routes. Instead, it is a scheduling problem more similar to the Nurse Scheduling Problem~\cite{Burke2004TheSO}, where a series of tasks have to be scheduled over a time interval so that they meet certain constraints. In addition, the Cash Management problem has an associated optimization problem since the cash delivery to each branch or ATM has an associated cost, which depends generally on its location.

Therefore, the Cash Management optimization problem consists of finding the optimal scheduling of cash delivery to a network of branches and ATMs in a given geography in a way that the cost of the transactions performed is minimized, while satisfying three requirements: 

\begin{itemize}
    \item On each day of the time interval considered, the amount of cash in each branch and ATM must be higher than a pre-established minimum that guarantees the operability of the branch or ATM, and less than a maximum for security reasons.
    
    \item We can send or withdraw money from the branch or ATM only once a day.
\end{itemize}
In addition, there are a number of hard constraints that must be taken into account in the optimization process: 
\begin{itemize}
    \item The number of deliveries made on a given day must not exceed a pre-established maximum due to the limited number of available delivery trucks.
    
    \item Shipments and withdrawals are made in discrete cash amounts, thus facilitating the management of the entire operation.
    
    \item Total cash in the branch and ATM network cannot exceed a pre-set maximum amount on the last three business days of each month for regulatory reasons. 
    
\end{itemize}
Finally, it should also be noted that making a cash delivery on the first day of the time interval considered costs more money than any other day, since it is equivalent to making an urgent request only twenty-four business hours in advance. Additional conditions could be added to the description of the problem, such as the existence of non-working days for some ATM during the period, but in this manuscript we will limit to the Cash Management problem as described in this section.

\subsection{Mathematical formulation}
\label{sec_CMformulation}

Assume $C$ branches and ATMs (both will also be denoted as \textit{cash points}) distributed according to a particular geography, and let's suppose that we want to plan the daily cash delivery at each of the $C$ cash points for a period of $D$ days. 
We will label each cash point and each day of the period with the integers $c\in\left[ 0,C\right)$ and $t\in\left[ 0,D\right)$ respectively. 
The CMP is completely defined by the following sets of variables,
\begin{itemize}
    \item $k^0_c\in\mathbb{R}^{+}$ : price of sending or withdrawing money at the branch or ATM $c$ on the first day of the period under consideration.
    
    \item $k_c \in \mathbb{R}^+, k_c < k^0_c$ : price of sending or withdrawing money at the branch or ATM $c$ on any day except the first day of the considered temporal period.
    
    \item $p_{ct}\in\mathbb{Z}$ : initial prediction of the cash that will be available at cash point $c$ on day $t$. Such a prediction may be outside the limits imposed on the amount of daily cash that each branch and ATM must have.
    
\end{itemize}
In addition to the variables that will define the constraints of the optimization problem:
\begin{itemize}
    \item $v_l,v_h\in \mathbb{R}^+$ : minimum and maximum available cash respectively that each branch and ATM must have each day of the interval to guarantee its operability and for security reasons.
    
    \item $v_f\in \mathbb{R}^+$ : maximum value of total cash in the network of branches and ATMs that can be held on the last days of the time interval under consideration. For simplicity, and without loss of generality, we will set the last day of the time interval as the day subject to this restriction ($t=D-1$).
    
    \item $l\in \mathbb{N}$ : maximum number of transactions (shipments and withdrawals) that can be made throughout the branch and ATM network each day.
    
\end{itemize}
The value of these variables cannot be modified in the optimization process, and will be given by the particularities of each geographical area and time interval. From these fixed variables it is possible to code the CMP with different cost functions by changing the definition of the variables to be optimized. 

In the present formulation, the variables to be optimized are the discretized cash available at each branch and ATM $c$ each day $t$, $m_{ct}$. For convenience, we define the normalized cash $M_{ct}$
\begin{equation}
    M_{ct} = \dfrac{m_{ct} - v_l}{\Delta m} = n\,,
\end{equation}
where $n = 0, 1,...,h$, $m_{ct} = v_l+n\Delta m$, $\Delta m = (v_h-v_l)/(h-1)$ is the spacing between levels, and $h$ is the number of discrete cash values allowed. Note that a total of $V = \log_2(h)\cdot C\cdot D$ binary, $x=\{0,1\}$, or spins, $z=(2x-1)$, variables will be necessary for the encoding. For simplicity, and without loss of generality, we take $h=4$, $v_h=h-1$, and $v_l=0$, so that 
\begin{equation}
M_{ct} = 3/2 + (1/2) z_{ct}^0 + z_{ct}^1 \in \{0,1,2,3\} \,\,,
\end{equation} 
where $z_{ct}^i\in\{-1,+1\}$ with $i=0,1$. The advantage of this formulation is that we impose that all quantities are within the bounds set by the constraints of the problem, i.e., by definition it is satisfied that $M_{ct}\in\left[v_l,v_h\right]$.

We further note that money transfers and withdrawals are always made with cash packages of discrete amounts multiple of $\Delta m$ as stated in the problem description. This fact makes that the initial prediction is also discretized $p_{ct}\in\mathbb{Z}$ by $\Delta m$, but can take an infinite number of values unlike $M_{ct}$ (see Fig.~\ref{fig_singleinstanceCMP}c). Note that $p_{ct}$ is a non-exact estimate, so approximating its value to the nearest allowed integer does not imply any noticeable shortcoming in practical application.

To construct the cost function of the problem it is convenient to define the cash that would be available at the cash point $c$ on the day $t$ if no shipment or withdrawal is made at that location on that day,
\begin{equation}
    \begin{split}
    & W_{ct} = p_{ct}+(M_{c,t-1}-p_{c,t-1}) \,\,\, \forall \, d\neq 0 , \\
    & W_{c0} = p_{c0} .
    \end{split}
\end{equation}
Therefore, the cost function to be minimized in order to optimize the cost of shipments and withdrawals is as follows
\begin{dmath}
    C(z) = \sum_{c\in[0,C)}k_c^0 \left[1-\delta\left(M_{c0} - p_{c0}\right)\right] + \sum_{c\in[0,C),t\in[1,D)}k_c\left[1-\delta\left(M_{ct} - W_{ct}\right)\right],
\end{dmath}
where $\delta(x)=1$ if $x=0$, and $\delta(x)=0$ otherwise. Note that solving this optimization problem is equivalent to finding the ground state of the following Hamiltonian:
\begin{widetext}
\begin{equation}
\begin{split}
    \hat{C}(\hat{Z})= & \sum_{c\in[0,C)}k_c^0 \left[1-\delta\left(\frac{3}{2}+\frac{1}{2}\hat{Z}_{c0}^0 + \hat{Z}_{c0}^1 - p_{c0}\right)\right] + \\ & \sum_{c\in[0,C),t\in[1,D)}k_c\left[1-\delta\left(\frac{1}{2}\hat{Z}_{ct}^0 + \hat{Z}_{ct}^1-\frac{1}{2}\hat{Z}_{c,t-1}^0 - \hat{Z}_{c,t-1}^1 - p_{ct}+ p_{c,t-1}\right)\right],
\end{split}
\end{equation}
\end{widetext}
where $\hat{Z}_{ct}^i$ denotes the Pauli Z operator acting on the corresponding qubit. Notice that the above Hamiltonian is defined exclusively in terms of Pauli Z operators such that their eigenstates are separable and computational basis states.

As discussed in the previous section, the optimization of the Cash Management problem is also subject to constraints which must be formulated in terms of inequalities:
\begin{itemize}
    \item On the last day of the time interval the sum of cash from all cash points in the network cannot exceed a certain amount:
    \begin{equation}
        G_D \equiv \sum_{c\in[0,C)}M_{cD} \leq v_f\,\,.
    \label{eq_cashnetworkcons}
    \end{equation}
    
    \item There is a limit to the daily number of shipments and withdrawals made throughout the network: 
    \begin{equation}
        N_t \equiv \sum_{c\in[0,C)}\left[1-\delta\left(M_{ct} - W_{ct}\right)\right] \leq l \,\,.
    \end{equation}
\end{itemize}
Therefore, there are a total of $D+1$ additional constraints. One way to include these constraints in the optimization process is to include penalty terms in the cost function that raise its value when a solution fails to satisfy any of the constraints. In this manuscript, we explore the advantages of performing a multi-objective optimization of the cost function that takes into account these constraints without the need to include additional terms. 


\subsection{Single instance example}

\begin{figure}
  \centering
\includegraphics[width=0.985\linewidth]{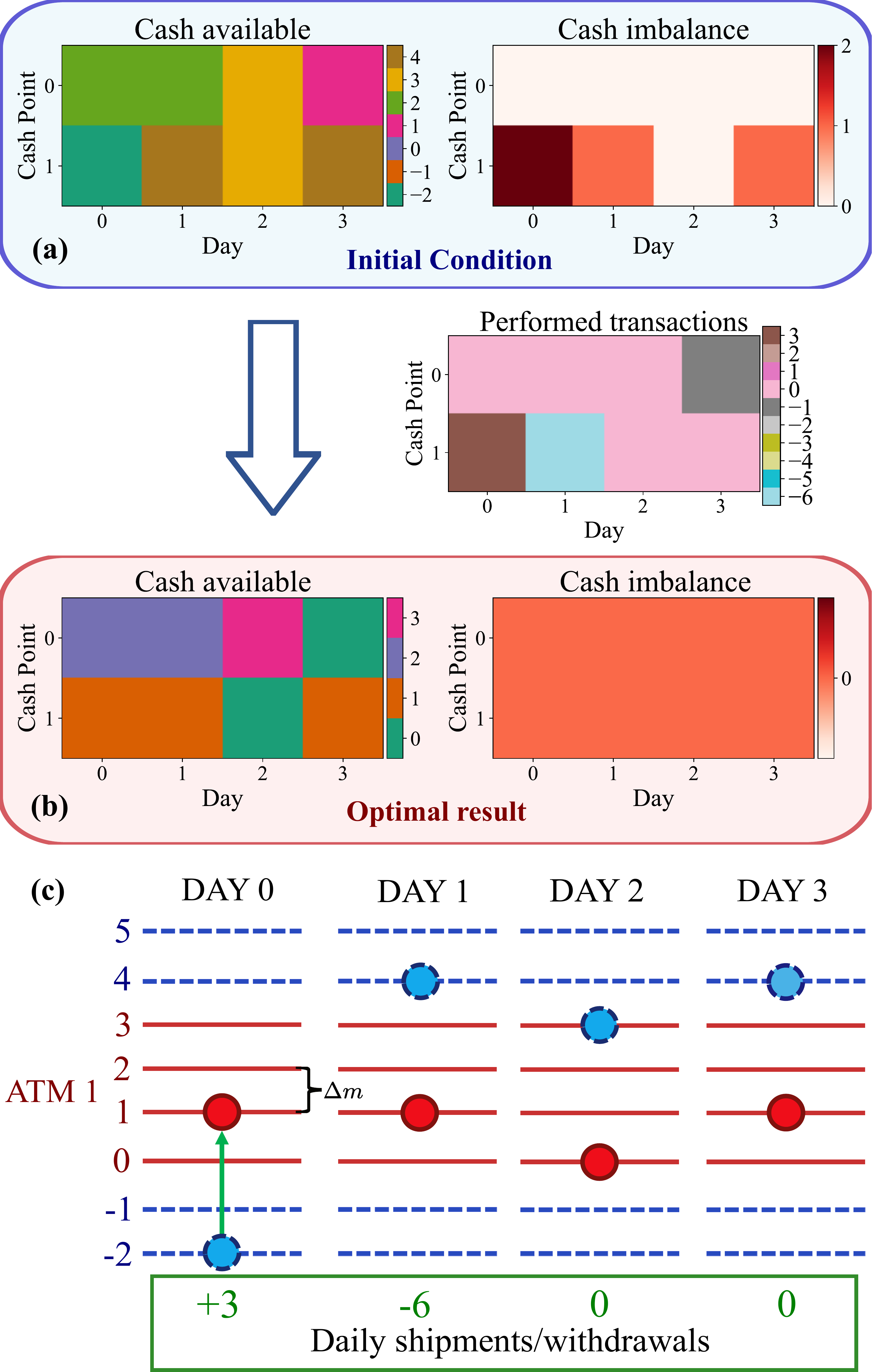}
\caption{Example of a single instance of a CMP with two cash points optimized over a four-day period. (a) Situation given by the initial prediction $p_{cd}$ when no transactions are performed. (b) Optimal solution $m_{cd}$. (c) Scenario for cash point one in which the red dashed and solid blue lines show the discrete cash levels inside and outside the constraint interval $\left[v_l,v_h\right]$ respectively, while the red dashed and solid blue circles show the initial prediction and the optimal result respectively (Color online).}
  \label{fig_singleinstanceCMP}
\end{figure}

For the sake of clarity, we include an example of a CMP with two cash points where the cash transactions are optimized during a period of four days. The practical scenario is shown in Fig.~\ref{fig_CMPscheme}b. In the discretized space the initial prediction of the available cash in each branch or ATM is
\begin{equation*}
    p_{ct} = \begin{pmatrix}
 2 & 2 & 3 & 1 \\ 
 -2 & 4 & 3 & 4 
\end{pmatrix} ,
\end{equation*}
with delivery costs $k_0^0 = 4 , k_1^0 = 8$, and $k_0 = 2, k_1 = 4$. The constraints of the problem are defined by $v_l=0,v_h=3,v_f=1,l=1$. 
As can be seen in Fig.~\ref{fig_singleinstanceCMP}a, without any transaction, cash point 1 violates the constraint imposed on available cash. The constraint on the total cash value, as the sum of the cash from the two points, is also violated on the last day of the interval. The optimal state of this instance, which satisfies all the constraints and minimizes the delivery costs, will be
\begin{equation*}
    M_{ct} = \begin{pmatrix}
 2 & 2 & 3 & 0 \\ 
  1 & 1 & 0 & 1 
\end{pmatrix} ,
\end{equation*}
or in terms of the spin variables $z^i_{ct}$,
\begin{equation*}
    M_{ct} = \scriptscriptstyle{\begin{pmatrix}
 (-1,1)&(-1,1)&(1,1)& (-1,-1) \\ 
  (1,-1)&(1,-1)&(-1,-1)&(1,-1)  
\end{pmatrix}} ,
\end{equation*}
that corresponds to performing the transactions (shipments and withdrawals) shown in Fig.~\ref{fig_CMPscheme}b and Fig.~\ref{fig_singleinstanceCMP}. The cost $C$ of the optimal solution is 10. 

\section{Numerical results}
\label{sec_numresults}

\begin{figure*}
  \centering
\includegraphics[width=\linewidth]{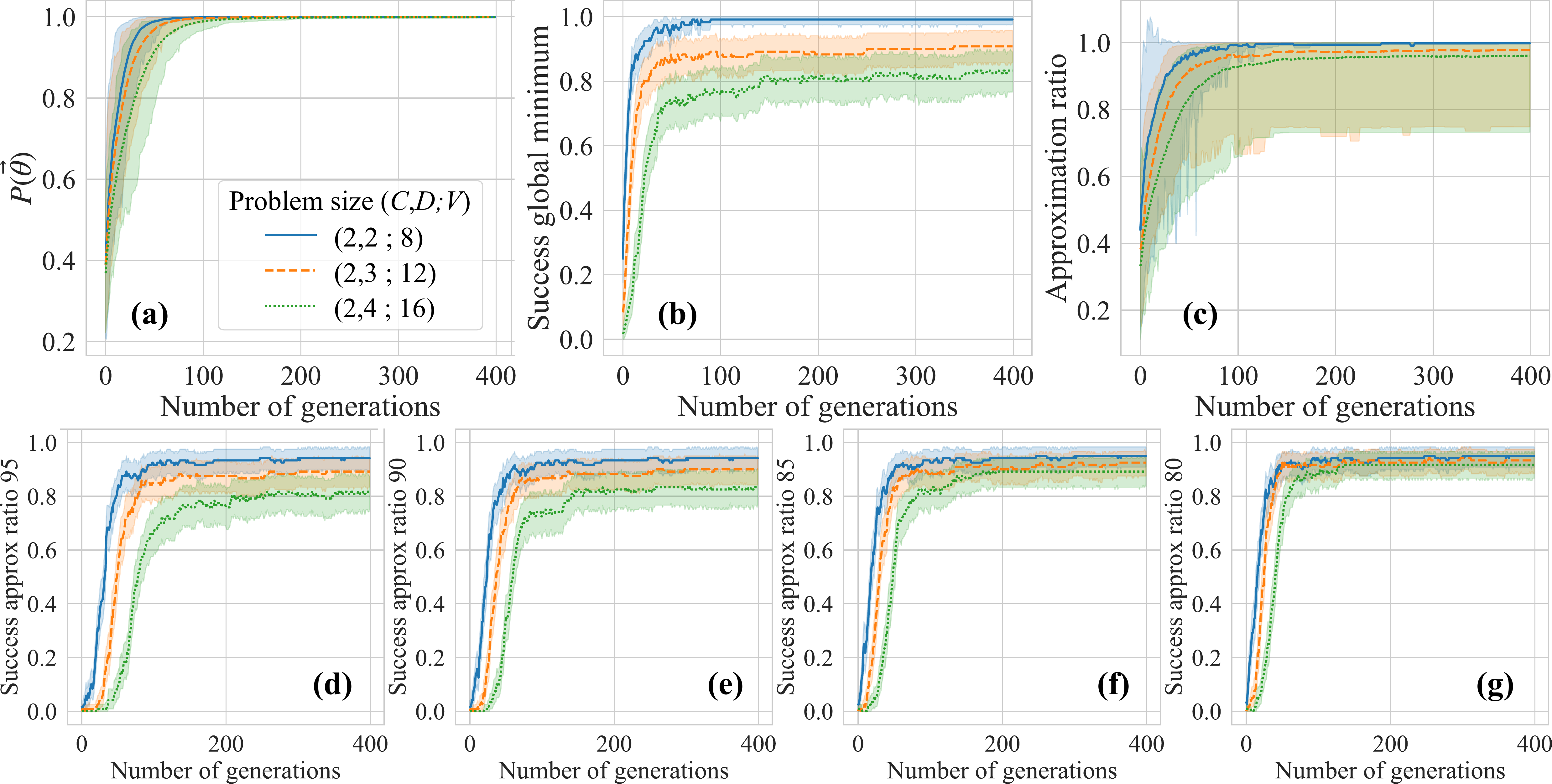}
\caption{Empirical results from the resolution of 120 CMP instances (2 cash points within a period of 2,3, and 4 days; $V=2CD$ is the number of variables/qubits required for encoding the problem) with the classical simulation of MOVCO-VQE using the single-layer ansatz~\eqref{eq_ryansatz}. (a) Average of the percentage of constraints that are satisfied by the 8192 sampled solutions. (b) Percentage of instances where the variational wavefunction reached an overlap with the global minimum higher than 0.1. (c) Average of the approximation ratio. (d-g) Success rates in which an instance is successful if it achieves an approximation ratio higher than 0.95,0.9,0.85, and 0.8 respectively. Plots (a) and (c) show a $95\%$ percentile interval around the average ranging from 2.5 to 97.5 percentiles of the distribution. The plots (b) and (d-g) display the $95\%$ confidence interval in the average estimation (Color online).}
  \label{fig_MOVQE_results}
\end{figure*}

\begin{figure}
  \centering
\includegraphics[width=\linewidth]{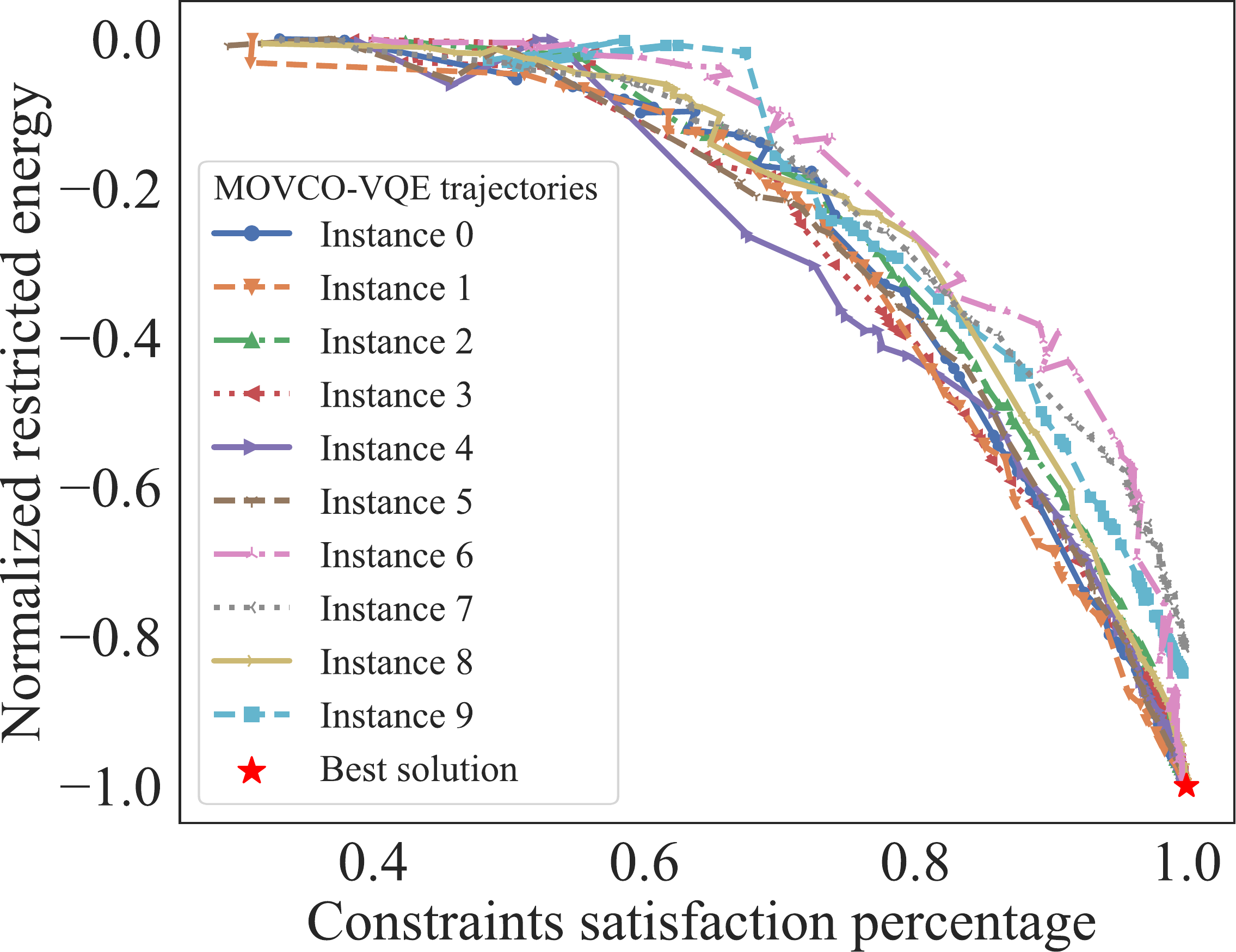}
\caption{Evolution of the solution reached by MOVCO-VQE when solving 10 independent CMP instances with 2 cash points over a period of 4 days. Each line corresponds to the trajectory of an instance in the fitness functions space as we increase the number of iterations of the algorithm. The restricted energy is normalized between 0 and -1 for all samples  (Color online).}
  \label{fig_MOVQEtrajectories}
\end{figure}

Next, we test the proposed variational optimization method on a real-world problem of great industrial interest and challenging constraints: the Cash Management problem formulated in the previous section. Due to the characteristics of this problem, we use the VQE variant of the algorithm (MOVCO-VQE) because of its flexibility in constructing shallow ansätze with high expressibility. In addition, we shall make a comparison with the results obtained by optimizing the CMP using a Variational Quantum Eigensolver where the constraints are encoded as penalty terms. The computational experiments have been performed on CMP instances randomly generated as explained in Appendix~\ref{append_problemgeneration}.

\subsection{Performance of the multiobjective approach}
\label{sec_numresultsMOVCO}

We analyze the convergence and the efficiency of the algorithm as we increase the number of generations for different problem sizes (Fig.\ref{fig_MOVQE_results}). We use one layer of the VQE ansatz described in section~\ref{sec_vqe} where the variational parameters are updated following the NSGA-II routine with binary tournament selection, simulated binary crossover \cite{Deb1995SimulatedBC}, and polynomial mutation \cite{Deb1996ACG} as genetic operators. The population of the genetic algorithm was set to 10, also creating an offspring of ten new individuals in each generation. This means that the variational algorithm evaluates the quantum circuit with a different set of parameters $\vec{\theta}$ ten times per generation. Here and in the subsequent results, the quantum circuits are classically simulated under noise-free conditions, sampling $K=8192$ bit strings from the final probability distribution that mimic the finite number of measurements performed on a quantum computer.

To conduct the benchmark we will use metrics that tell us about the quality of the solutions in terms of the number of constraints that they satisfy and the transactions cost of these solutions, regardless of the VQA we apply. As for the constraint satisfaction, we may straightforwardly use the fitness function $P(\vec{\theta}_{sol})$ from Eq.~\eqref{eq_fitnessprojection}, where $\vec{\theta}_{sol}$ are the final variational parameters. Regarding the transactions cost we shall use the approximation ratio $\epsilon(\vec{\theta})$ defined as
\begin{equation}
    \epsilon(\vec{\theta}) = \dfrac{C_{max} - \bra{\Psi (\vec{\theta})}\hat{C}\ket{\Psi (\vec{\theta})}}{C_{max} - C_{min}}\,,
\end{equation}
where $C_{max}=\sum_c k^0_c + (D-1)\sum_c k_c$ is the maximum possible cost, and $C_{min}$ is the minimum achievable cost provided that all constraints are satisfied. Hence, when $P(\vec{\theta}_{sol})=1$ and $\epsilon(\vec{\theta}_{sol})=1$ the algorithm achieved the best solution, i.e the global minimum that fulfills the constraints $\ket{\Psi_{gs}}$. We also analyze the overlap of the variational wavefunction with the global minimum $\rho = |\langle\Psi (\vec{\theta}) | \Psi_{gs} \rangle|^2$ by the success rate. We consider that the global minimum is easily sampled from the wavefunction, and thus has achieved success, if $\rho>0.1$. 

In Fig.\ref{fig_MOVQE_results}a, we observe how the method prevents the variational wave function from being trapped in local minima outside the feasible space. Indeed, after two hundred generations of the algorithm the solutions of all instances satisfy all the constraints of the CMP. Moreover, these solutions have low transaction costs, as shown in Fig.\ref{fig_MOVQE_results}c-g. Although  Fig.\ref{fig_MOVQE_results}c reflects some variability in the approximation ratio, the mean is high indicating that most instances achieve low energy solutions. This fact can also be seen in Fig.\ref{fig_MOVQE_results}d-g where we plot the percentage of instances that achieved an approximation ratio $\epsilon(\vec{\theta}_{sol})>0.95,0.9,0.85,0.8$ respectively. Even for the largest problems, most instances reach a solution whose approximation ratio is higher than $80\%$. As for the exact optimum, we see in Fig.\ref{fig_MOVQE_results}b that that for about $80\%$ of the instances with 2 cash points over a period of 4 days, and about $90\%$ of the instances with 2 cash points over a period of 3 days, the variational wave function allows to efficiently sample the global minimum after only a hundred generations.

We may divide the execution process of the algorithm into two phases (see Fig.~\ref{fig_MOVQEtrajectories}). During the first stage, the training raises the projection $P(\vec{\theta})$ while the energy $E(\vec{\theta})$ might stay steady at 0. When the projection is high enough, the wavefunction starts sampling solutions in $\mathcal{S}$, and the energy is simultaneously minimized as the projection keeps increasing so that progressively the optimization is performed only on the feasible subspace. This behavior can not only accelerate convergence in the second stage, but also avoids getting stuck in local minima and greatly reduces the probability of ending up outside the in-constraint subspace thanks to the training performed in the first stage. For small-size systems or problems with a high density of states satisfying all constraints, such as those implemented in Fig.\ref{fig_MOVQE_results} and Fig.\ref{fig_MOVQEvsVQE_results}, the first stage can be quickly overcome. In this scenario the energy is also minimized from the first generations. 

\subsection{Comparison with the penalties approach}
\label{sec_MOVCOVSpenVQE}

\begin{figure*}
  \centering
\includegraphics[width=0.935\linewidth]{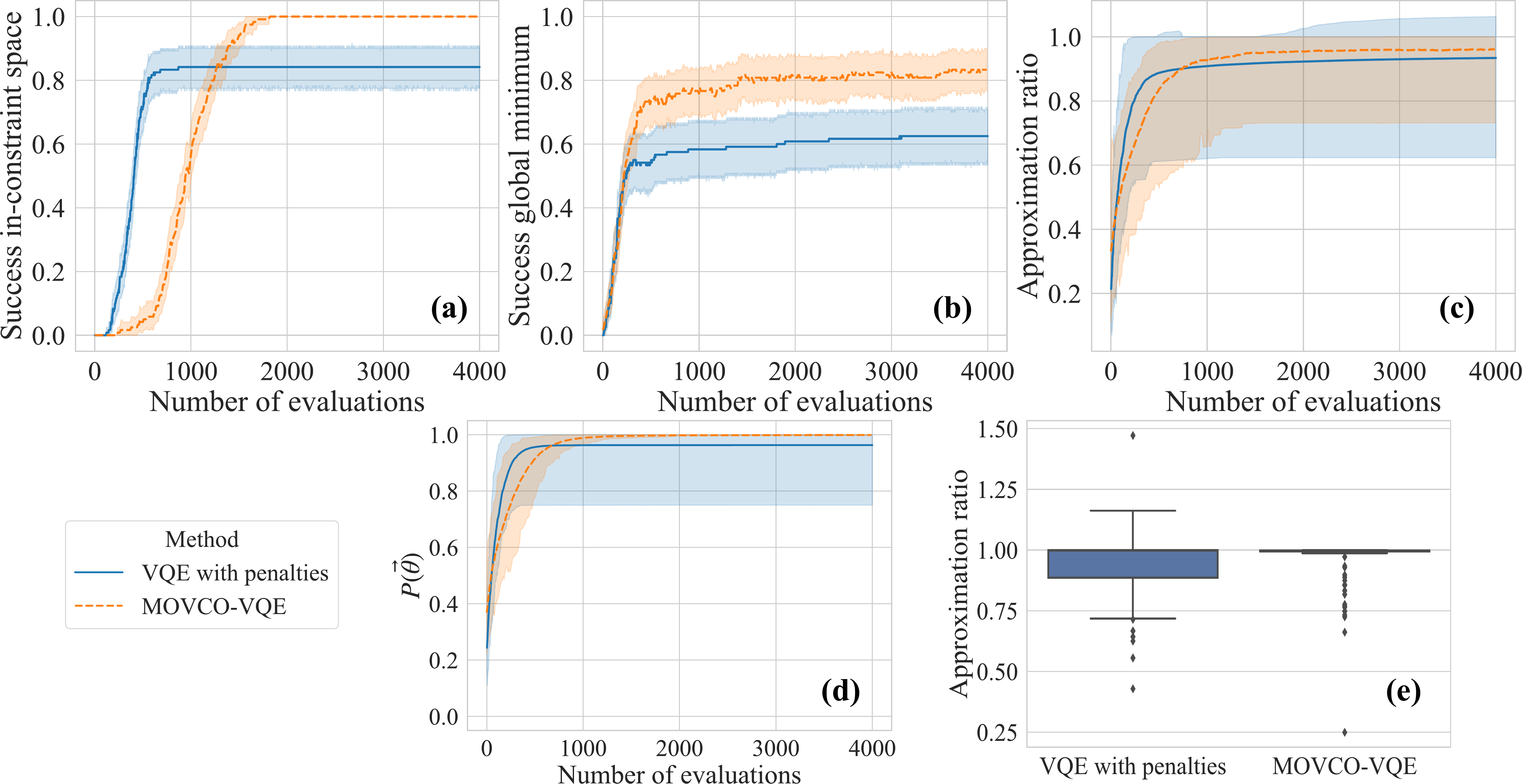}
\caption{Comparison between the MOVCO-VQE and VQE with penalties performance. We show the results from 120 CMP instances (2 cash points within a period of 4 days). We conducted a classical simulation of both algorithms using the single-layer ansatz~\eqref{eq_ryansatz} and 8192 shots. (a) Percentage of instances whose almost all sampled solutions are within the feasible space. (b) Percentage of instances where the variational wavefunction reached an overlap with the global minimum higher than $0.1$. (c,e) Average and distribution respectively of the approximation ratios. (d) Average of the percentage of constraints that are satisfied by the sampled solutions. Plots (c) and (d) show a $95\%$ percentile interval around the average ranging from 2.5 to 97.5 percentiles of the distribution. Plots (a) and (b) displays the $95\%$ confidence interval in the average estimation. The boxes in plot (e) show the lower and the upper quartiles, and the whiskers extend to 1.5 interquartile ranges (Color online).}
  \label{fig_MOVQEvsVQE_results}
\end{figure*}

After confirming the good convergence to optimal solutions of the multiobjective approach, we study the advantages of MOVCO over a standard approach. 

A traditional technique to deal with hard constraints is transforming the original Hamiltonian including \textit{penalty terms}. These terms increase the expectation value of the Hamiltonian when the solutions do not fulfill the constraints. In the Cash Management problem, the penalized Hamiltonian may be expressed as   
\begin{equation}
\begin{split}
    \hat{C}_{pen}(\hat{Z})= \hat{C}(\hat{Z}) & + \lambda_f L\left(G_D - v_f\right)\\ & +  \lambda_l \sum_{t\in[0,D)} L\left(N_t - l\right)        
\end{split}
\label{eq_penham}
\end{equation}
where $\hat{C}$, $G_D$, and $N_t$ are defined in section~\ref{sec_CMformulation}, $L$ is the Heaviside step function ($L(x)=1$ if $x\geq0$ , $L(x)=0$ otherwise, $x\in\mathbb{R}$), and $\lambda_{f}$ and $\lambda_l$ are tunable positive hyperparameters. Thereby, the variational optimization becomes the search of the state that minimizes $\bra{\Psi(\vec\theta)}\hat{C}_{pen}(\hat{Z})\ket{\Psi(\vec\theta)}$. Note that this state will satisfy all constraints if $\lambda_{f}$ and $\lambda_l$ are large enough.  

We compare MOVCO-VQE with the VQE with penalties in CMP instances with 2 cash points optimized in a 4 days period, i.e 16-qubit problems. In VQE, the variational parameters are optimized by the simultaneous perturbation stochastic approximation (SPSA)~\cite{Spall_1992}. This gradient-free classical optimizer only performs two evaluations of the quantum circuit per iteration independently of the number of free parameters, making it one of the efficient methods for VQAs.~\cite{Cerezo_2021}. The value of the penalty hyperparameters was chosen in such a way as to ensure that all solutions that fail to satisfy any constraint of the problem have a higher cost than solutions in the feasible subspace. To improve the significance of the experiments, we tuned these values to maximize the average constraint satisfaction and the average approximation ratio as explained in Appendix~\ref{append_penaltiestunning}, setting $\lambda_{f}=\lambda_{l}=25$. For both MOVCO-VQE and VQE we use again the previously exposed one layer ansatz~\eqref{eq_ryansatz}, and the noise-free quantum computer simulation with $K=8192$ shots.  

As shown in Fig.\ref{fig_qinspired_MOVQEvsVQE_results}, MOVCO-VQE overcomes the algorithm with penalties in every considered metric. To make a fair comparison, we take into account that SPSA only needs two queries to the quantum processor per iteration, while MOVCO uses as much as the population size of the genetic algorithm (10 in our setup). Therefore, we plot the results in terms of the number of evaluations performed on the quantum computer. The main improvement is the ability of MOVCO-VQE to explore only the feasible space after a number of iterations, while VQE with penalties gets trapped in local minima outside the inconstraint regime. This fact can be seen in Fig.\ref{fig_MOVQEvsVQE_results}a, where we display the percentage of instances in which the overlap between the final wavefunction and the inconstraint subspace was almost total, i.e $P(\vec{\theta}_{sol})>0.99$. This behavior is also noticeable in Fig.\ref{fig_MOVQEvsVQE_results}d, where we directly show $P(\vec{\theta})$.

The percentage of instances achieving the global minimum is also increased and the cost of the solutions is reduced. Moreover, this improvement in convergence is associated with a lower dispersion of the results as seen in  Figs.\ref{fig_MOVQEvsVQE_results}c-e. In Fig.\ref{fig_MOVQEvsVQE_results}e the instances with an approximation ratio higher than 1 reached low energy but non-feasible solutions. One point to note is that MOVCO allows parallel processing of each of the quantum circuit evaluations performed in an iteration thanks to the genetic algorithm mechanism, which greatly reduces the execution time.

The advantage of MOVCO arises from the combination of the multiobjective approach and the genetic algorithm technique. As a test, we also analyzed the performance of VQE with penalties when optimizing the variational parameter using a single objective genetic algorithm. Specifically, we used the same genetic operators and hyperparameters as for MOVCO, but the ordering of the solutions performed in each iteration was done based on the average value of the penalized energy $\bra{\Psi(\vec\theta)}\hat{C}_{pen}(\hat{Z})\ket{\Psi(\vec\theta)}$. As shown in Appendix \ref{append_singleobjectiveGAoptimization}, the result is similar to that obtained with SPSA in Fig.\ref{fig_qinspired_MOVQEvsVQE_results} thus the advantage of MOVCO is maintained by the proposed multi-objective optimization. 

\subsection{Benchmarking for larger systems by product states}

\begin{figure}[t]
  \centering
\includegraphics[width=\linewidth]{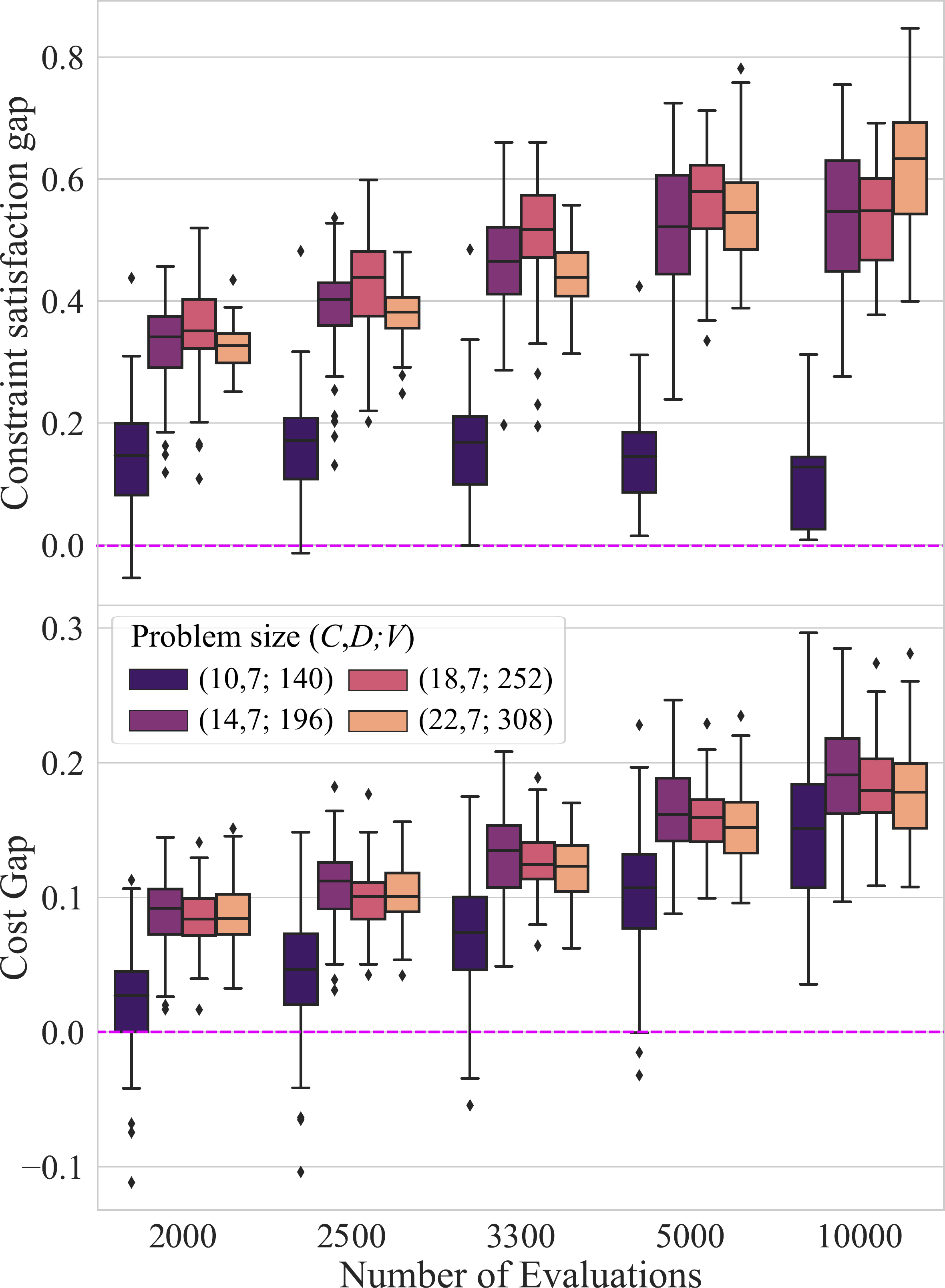}
\caption{Comparison between the performance of the MOVCO approach and the penalty approach using product states, as we increase the number of iterations. The same 80 instances are solved by both algorithms, and we compare the quality of the solutions by (a) the difference in the percentage of constraints that are satisfied by the sampled states, and (b) the percentage difference between the transactions cost of the sampled solutions. In cases above the pink dashed line, MOVCO achieved a higher quality solution. The boxes show the lower and the upper quartiles, the whiskers extend to 1.5 interquartile ranges, and the remaining points are observational data outside this distribution (Color online).}
  \label{fig_qinspired_MOVQEvsVQE_results}
\end{figure}

The classical simulation of entangled states such as~(\ref{eq_ryansatz}) involves an exponential cost of resources which prevents the analysis of problems with a high number of variables. In order to increase the number of cash points and days in our computational experiments, and compare how the performance of both algorithms evolves in this scenario, we need to reduce the complexity of the variational ansatz. For this reason, we now analyze the performance of the two methods using the fully separable ansatz,
\begin{equation}
  \label{eq_product_state}
  \ket{\Psi_{sp} (\vec{\theta})} = \prod_{n}^{N}e^{i\theta_{n}\hat{Y}_n} \ket{0} = \bigotimes_{n=1}^N \ket{\psi (\theta_n)},
\end{equation}
with
 $\ket{\psi (\theta_n)}  = \cos{\theta_n} \ket{0} + \sin{\theta_n} \ket{1}$. This variational form is a borderline case of the ansatz~(\ref{eq_ryansatz}) with 0 layers. Since we eliminate the entanglement of the quantum circuit, the variational wavefunction probability distribution of~(\ref{eq_product_state}) can be efficiently sampled using an array of just $N$ elements $\{\cos^2{\theta_n}\}$. Note that this simple architecture is already able to reproduce any classical state from the Hilbert space. Therefore, there always exist a set of parameters $\vec{\theta}_{sol}$ for which~(\ref{eq_product_state}) is the ground state of the CMP Hamiltonian. The use of this family of states may be seen as a quantum-inspired alternative to the variational quantum algorithms with entangling layers tested in the previous results.  

We test the algorithms by optimizing ATM networks consisting of 10 to 22 cash points for a whole week, which corresponds to problems up to $V=308$ variables. Due to the increased size of the problems, in these results we apply MOVCO-VQE with a population set to 100, and VQE with penalties where $\lambda_{f}=\lambda_{l}=50$. The number of bit strings sampled per evaluation is again K = 8192. The procedure is as follows. We generate one random instance of the CMP as previously explained, and resolve the optimization problem by both algorithms. The larger size of these instances does not enable us to know the exact solution to the problem by exhaustive search, so we performed the benchmarking between the two methods by directly comparing the solutions obtained. We use two metrics: the constraint satisfaction gap defined as
\begin{equation}
    P_{gap} = P(\vec{\theta}_{MOVCO}) - P(\vec{\theta}_{VQEpen})  
\end{equation}
and the cost gap
\begin{equation}
    C_{gap} = \dfrac{C_{VQEpen} - C_{MOVCO} }{C_{VQEpen}}
\end{equation}
where $C_{i} = \bra{\Psi_{sp} (\vec{\theta}_{i})}\hat{C}\ket{\Psi_{sp} (\vec{\theta}_{i})}$ with $i=\{VQEpen,MOVCO\}$, and $\vec{\theta}_{MOVCO},\vec{\theta}_{VQEpen}$ the final angles after the MOVCO-VQE and VQE with penalties algorithms respectively. 

In Fig.\ref{fig_qinspired_MOVQEvsVQE_results} we display the evolution of the solutions as we increase the number of cost function evaluations. The cases above the pink line reflect better solutions for MOVCO-VQE. The improvement is noticeable in both constraint satisfaction and transaction cost of the sampled solutions and becomes more pronounced as we increase the iterations of the algorithms. Indeed, we achieve better solutions for $100\%$ of the instances when the algorithms have performed 10000 evaluations.     

\section{Conclusions}

In this manuscript, we introduce the Multi-Objective Variational Constrained Optimizer (MOVCO), a variational quantum method to solve combinatorial optimization problems (CO) with hard constraints. This method allows improving the performance of Variational Quantum Algorithms (VQAs) through a genetic algorithm that simultaneously optimizes two fitness functions: one dealing with the satisfaction of the constraints, and the other with the energy minimization of the solutions within the feasible subspace. 

We provide empirical evidence of the robust performance of MOVCO on a very relevant industrial problem, the Cash Management problem (CMP). We propose a novel formulation of CMP in terms of binary variables and solve it using the MOVCO version of VQE. We compare these results with a standard approach in which VQE optimizes a cost function with penalty terms that artificially increase the energy of infeasible states. Our study reveals that MOVCO provides benefits in avoiding unfeasible minima while enhancing convergence to lower energy solutions. A detailed study of the influence of the hyperparameters of the genetic algorithm on MOVCO performance, such as population size or genetic operators, may lead to better results.   

The MOVCO method is ansatz agnostic, so it can be applied to a wide range of VQAs beyond VQE. New studies may be conducted implementing the method on problem-dependent ansätze such as QAOA where the CO, usually formulated as a QUBO, is mapped to the quantum processor. Further enhancements can be incorporated into MOVCO. For example, a CVaR version of MOVCO can be analyzed, in which only the lowest energy percentage of the sampled states would be used to compute the  fitness functions of the multiobjective optimization. Since CO does not require a complete overlap between the variational wavefunction and the in-constraint subspace but a high probability of sampling low-energy feasible solutions, another interesting possibility is to set an upper limit to the percentage of sampled solutions that satisfy all constraints.          

This work provides further insight into the application of variational algorithms to optimization problems of practical interest. Furthermore, this is, to the best of our knowledge, the first attempt to solve CMP with quantum computing. As such, this work may open the way for further studies on this and other problems in the large set of real-world combinatorial optimization problems with hard constraints. 

\begin{acknowledgments}
 This work was supported by the Spanish CDTI through Misiones Ciencia e Innovación Program (CUCO) under Grant MIG-20211005, PID2021-127968NB-I00 funded by MCIN/AEI/10.13039/501100011033/FEDER,UE, and CSIC Interdisciplinary Thematic Platform (PTI) Quantum Technologies (PTI-QTEP). P. D.-V. also acknowledges support from CAM/FEDER project No. S2018/TCS-4342 (QUITEMAD-CM). The authors also gratefully acknowledge the Scientific computing Area (AIC), SGAI-CSIC, for their assistance while using the DRAGO Supercomputer for performing the simulations, and Centro de Supercomputación de Galicia (CESGA) who provided access to the supercomputer FinisTerrae. 
\end{acknowledgments}

\section*{Author Contributions Statement}
P. D.-V. developed the mathematical formulation of the optimization problem, designed the quantum algorithm and performed the numerical calculations.  J. L.-H. and S. H.-S. formulated the cash-management problem. D. P., E. S.-M. and J. J. G.-R. conceived and supervised the research, all authors contributed to writing the manuscript.

\section*{Disclaimer}

This paper is purely scientific and informative in nature and is not a product of BBVA SA or any of its subsidiaries. Neither BBVA nor such subsidiaries are aware of or necessarily share the premises, conclusions or contents in general of this document. Consequently, the responsibility for its originality, accuracy, reliability or for any other reason lies exclusively with the authors. This document is not intended as investment research or investment advice, or a recommendation, offer or solicitation for the purchase or sale of any security, financial instrument, financial product or service, or to be used in any way for evaluating the merits of participating in any transaction.

\bibliographystyle{unsrt}
\bibliography{bibliography}

\begin{thebibliography}{10}

\bibitem{Reiher_2017}
Markus Reiher, Nathan Wiebe, Krysta~M. Svore, Dave Wecker, and Matthias Troyer.
\newblock Elucidating reaction mechanisms on quantum computers.
\newblock {\em Proceedings of the National Academy of Sciences},
  114(29):7555--7560, jul 2017.

\bibitem{von_Burg_2021}
Vera von Burg, Guang~Hao Low, Thomas Häner, Damian~S. Steiger, Markus Reiher,
  Martin Roetteler, and Matthias Troyer.
\newblock Quantum computing enhanced computational catalysis.
\newblock {\em Physical Review Research}, 3(3), jul 2021.

\bibitem{Lubasch_2020}
Michael Lubasch, Jaewoo Joo, Pierre Moinier, Martin Kiffner, and Dieter Jaksch.
\newblock Variational quantum algorithms for nonlinear problems.
\newblock {\em Physical Review A}, 101(1), jan 2020.

\bibitem{Garc_a_Molina_2022}
Paula Garc{\'{\i} }a-Molina, Javier Rodr{\'{\i}}guez-Mediavilla, and
  Juan~Jos{\'{e}} Garc{\'{\i}}a-Ripoll.
\newblock Quantum fourier analysis for multivariate functions and applications
  to a class of schrödinger-type partial differential equations.
\newblock {\em Physical Review A}, 105(1), jan 2022.

\bibitem{Biamonte_2017}
Jacob Biamonte, Peter Wittek, Nicola Pancotti, Patrick Rebentrost, Nathan
  Wiebe, and Seth Lloyd.
\newblock Quantum machine learning.
\newblock {\em Nature}, 549(7671):195--202, sep 2017.

\bibitem{Benedetti_2019}
Marcello Benedetti, Erika Lloyd, Stefan Sack, and Mattia Fiorentini.
\newblock Parameterized quantum circuits as machine learning models.
\newblock {\em Quantum Science and Technology}, 4(4):043001, nov 2019.

\bibitem{Perdomo-Ortiz_2018}
Alejandro Perdomo-Ortiz, Marcello Benedetti, John Realpe-Gómez, and Rupak
  Biswas.
\newblock Opportunities and challenges for quantum-assisted machine learning in
  near-term quantum computers.
\newblock {\em Quantum Science and Technology}, 3(3):030502, jun 2018.

\bibitem{Nikolaj2018}
Nikolaj~Moll et~al.
\newblock Quantum optimization using varational algorithms on near-term quantum
  devices.
\newblock {\em Quantum Sci. Technol}, 3 030503, 2018.

\bibitem{Nemhauser1988IntegerAC}
George~L. Nemhauser and Laurence~A. Wolsey.
\newblock {\em Integer and Combinatorial Optimization}.
\newblock John Wiley \& Sons, Ltd, 1988.

\bibitem{Lenstra1975}
J.~K. Lenstra and A.~H. G.~Rinnooy Kan.
\newblock Some simple applications of the travelling salesman problem.
\newblock {\em Journal of the Operational Research Society}, 26(4):717--733,
  1975.

\bibitem{Goemans1995ImprovedAA}
Michel~X. Goemans and David~P. Williamson.
\newblock Improved approximation algorithms for maximum cut and satisfiability
  problems using semidefinite programming.
\newblock {\em J. ACM}, 42:1115--1145, 1995.

\bibitem{Festa2002}
P.~Festa, P.M. Pardalos, M.G.C. Resende, and C.C. Ribeiro.
\newblock Randomized heuristics for the max-cut problem.
\newblock {\em Optimization Methods and Software}, 17(6):1033--1058, 2002.

\bibitem{Cerezo_2021}
M.~Cerezo, Andrew Arrasmith, Ryan Babbush, Simon~C. Benjamin, Suguru Endo,
  Keisuke Fujii, Jarrod~R. McClean, Kosuke Mitarai, Xiao Yuan, Lukasz Cincio,
  and Patrick~J. Coles.
\newblock Variational quantum algorithms.
\newblock {\em Nature Reviews Physics}, 3(9):625--644, aug 2021.

\bibitem{Preskill_2018}
John Preskill.
\newblock Quantum computing in the {NISQ} era and beyond.
\newblock {\em Quantum}, 2:79, aug 2018.

\bibitem{Bharti_2022}
Kishor Bharti, Alba Cervera-Lierta, Thi~Ha Kyaw, Tobias Haug, Sumner
  Alperin-Lea, Abhinav Anand, Matthias Degroote, Hermanni Heimonen, Jakob~S.
  Kottmann, Tim Menke, Wai-Keong Mok, Sukin Sim, Leong-Chuan Kwek, and Al{\'{a}
  }n Aspuru-Guzik.
\newblock Noisy intermediate-scale quantum algorithms.
\newblock {\em Reviews of Modern Physics}, 94(1), feb 2022.

\bibitem{Dominguez2017}
Amparo Domínguez, Angel Juan, and Renatas Kizys.
\newblock A survey on financial applications of metaheuristics.
\newblock {\em ACM Computing Surveys}, 50:1--23, 04 2017.

\bibitem{Sbihi2010CombinatorialOA}
Abdelkader Sbihi and Richard~W. Eglese.
\newblock Combinatorial optimization and green logistics.
\newblock {\em Annals of Operations Research}, 175:159--175, 2010.

\bibitem{Bezerra2021}
Camila Bezerra, Lucas Carneiro, Elck Carvalho, Thiago Chagas, Lucas Carvalho,
  Ana Uetanabaro, Gervásio Da~Silva, Erik Galvão Paranhos~da Silva, and
  Andréa Costa.
\newblock Artificial intelligence as a combinatorial optimization strategy for
  cellulase production by trichoderma stromaticum am7 using peach-palm waste
  under solid-state fermentation.
\newblock {\em BioEnergy Research}, 14, 12 2021.

\bibitem{ESKANDARPOUR201511}
Majid Eskandarpour, Pierre Dejax, Joe Miemczyk, and Olivier Péton.
\newblock Sustainable supply chain network design: An optimization-oriented
  review.
\newblock {\em Omega}, 54:11--32, 2015.

\bibitem{Kennedy2008ApplicationOC}
J.~Phillip Kennedy, Lyndsey Williams, Thomas~M. Bridges, R~Nathan Daniels, Dave
  Weaver, and Craig~W. Lindsley.
\newblock Application of combinatorial chemistry science on modern drug
  discovery.
\newblock {\em Journal of combinatorial chemistry}, 10 3:345--54, 2008.

\bibitem{Hadfield2017QuantumAO}
Stuart Hadfield, Zhihui Wang, Eleanor~Gilbert Rieffel, Bryan O’Gorman, Davide
  Venturelli, and Rupak Biswas.
\newblock Quantum approximate optimization with hard and soft constraints.
\newblock {\em Proceedings of the Second International Workshop on Post Moores
  Era Supercomputing}, 2017.

\bibitem{Hadfield_2019}
Stuart Hadfield, Zhihui Wang, Bryan O{\textquotesingle}Gorman, Eleanor Rieffel,
  Davide Venturelli, and Rupak Biswas.
\newblock From the quantum approximate optimization algorithm to a quantum
  alternating operator ansatz.
\newblock {\em Algorithms}, 12(2):34, feb 2019.

\bibitem{niroula2022constrained}
Pradeep Niroula, Ruslan Shaydulin, Romina Yalovetzky, Pierre Minssen, Dylan
  Herman, Shaohan Hu, and Marco Pistoia.
\newblock Constrained quantum optimization for extractive summarization on a
  trapped-ion quantum computer.
\newblock {\em Scientific Reports}, 12(1):1--14, 2022.

\bibitem{lucas2014ising}
Andrew Lucas.
\newblock Ising formulations of many np problems.
\newblock {\em Frontiers in physics}, page~5, 2014.

\bibitem{Hao2022}
Tianyi Hao, Ruslan Shaydulin, Marco Pistoia, and Jeffrey Larson.
\newblock Exploiting in-constraint energy in constrained variational quantum
  optimization, 2022.

\bibitem{Nannicini2019}
Giacomo Nannicini.
\newblock Performance of hybrid quantum-classical variational heuristics for
  combinatorial optimization.
\newblock {\em Phys. Rev. E}, 99:013304, Jan 2019.

\bibitem{Willsch_2020}
Madita Willsch, Dennis Willsch, Fengping Jin, Hans~De Raedt, and Kristel
  Michielsen.
\newblock Benchmarking the quantum approximate optimization algorithm.
\newblock {\em Quantum Information Processing}, 19(7), jun 2020.

\bibitem{Robert_2021}
Anton Robert, Panagiotis~Kl. Barkoutsos, Stefan Woerner, and Ivano Tavernelli.
\newblock Resource-efficient quantum algorithm for protein folding.
\newblock {\em npj Quantum Information}, 7(1), feb 2021.

\bibitem{Amaro_2022}
David Amaro, Matthias Rosenkranz, Nathan Fitzpatrick, Koji Hirano, and Mattia
  Fiorentini.
\newblock A case study of variational quantum algorithms for a job shop
  scheduling problem.
\newblock {\em {EPJ} Quantum Technology}, 9(1), feb 2022.

\bibitem{Azad_2022}
Utkarsh Azad, Bikash~K. Behera, Emad~A. Ahmed, Prasanta~K. Panigrahi, and Ahmed
  Farouk.
\newblock Solving vehicle routing problem using quantum approximate
  optimization algorithm.
\newblock {\em {IEEE} Transactions on Intelligent Transportation Systems},
  pages 1--10, 2022.

\bibitem{Leontica2022}
Sebastian Leontica and David Amaro.
\newblock Quantum optimization with instantaneous quantum polynomial circuits,
  2022.

\bibitem{Crooks2018}
Gavin~E. Crooks.
\newblock Performance of the quantum approximate optimization algorithm on the
  maximum cut problem, 2018.

\bibitem{Streif2020}
Michael Streif and Martin Leib.
\newblock Forbidden subspaces for level-1 quantum approximate optimization
  algorithm and instantaneous quantum polynomial circuits.
\newblock {\em Phys. Rev. A}, 102:042416, Oct 2020.

\bibitem{Diez2022}
Pablo Díez-Valle, Diego Porras, and Juan~José García-Ripoll.
\newblock Qaoa pseudo-boltzmann states, 2022.

\bibitem{Farhi_2022}
Edward Farhi, Jeffrey Goldstone, Sam Gutmann, and Leo Zhou.
\newblock The quantum approximate optimization algorithm and the
  sherrington-kirkpatrick model at infinite size.
\newblock {\em Quantum}, 6:759, jul 2022.

\bibitem{Farhi_2014}
Edward Farhi, Jeffrey Goldstone, and Sam Gutmann.
\newblock A quantum approximate optimization algorithm, 2014.

\bibitem{Peruzzo_2014}
Alberto Peruzzo, Jarrod McClean, Peter Shadbolt, Man-Hong Yung, Xiao-Qi Zhou,
  Peter~J. Love, Al{\'{a}}n Aspuru-Guzik, and Jeremy~L. O'Brien.
\newblock A variational eigenvalue solver on a photonic quantum processor.
\newblock {\em Nature Communications}, 5(1), jul 2014.

\bibitem{Bravyi2020}
Sergey Bravyi, Alexander Kliesch, Robert Koenig, and Eugene Tang.
\newblock Obstacles to variational quantum optimization from symmetry
  protection.
\newblock {\em Phys. Rev. Lett.}, 125:260505, Dec 2020.

\bibitem{Amaro_2022_FVQE}
David Amaro, Carlo Modica, Matthias Rosenkranz, Mattia Fiorentini, Marcello
  Benedetti, and Michael Lubasch.
\newblock Filtering variational quantum algorithms for combinatorial
  optimization.
\newblock {\em Quantum Science and Technology}, 7(1):015021, jan 2022.

\bibitem{Liu_2022}
Xiaoyuan Liu, Anthony Angone, Ruslan Shaydulin, Ilya Safro, Yuri Alexeev, and
  Lukasz Cincio.
\newblock Layer {VQE}: A variational approach for combinatorial optimization on
  noisy quantum computers.
\newblock {\em {IEEE} Transactions on Quantum Engineering}, 3:1--20, 2022.

\bibitem{Zhu_2020}
Linghua Zhu, Ho~Lun Tang, George~S. Barron, F.~A. Calderon-Vargas, Nicholas~J.
  Mayhall, Edwin Barnes, and Sophia~E. Economou.
\newblock An adaptive quantum approximate optimization algorithm for solving
  combinatorial problems on a quantum computer, 2020.

\bibitem{McArdle_2019}
Sam McArdle, Tyson Jones, Suguru Endo, Ying Li, Simon~C. Benjamin, and Xiao
  Yuan.
\newblock Variational ansatz-based quantum simulation of imaginary time
  evolution.
\newblock {\em npj Quantum Information}, 5(1), sep 2019.

\bibitem{Lee_2018}
Joonho Lee, William~J. Huggins, Martin Head-Gordon, and K.~Birgitta Whaley.
\newblock Generalized unitary coupled cluster wave functions for quantum
  computation.
\newblock {\em Journal of Chemical Theory and Computation}, 15(1):311--324, nov
  2018.

\bibitem{Wecker_2015}
Dave Wecker, Matthew~B. Hastings, and Matthias Troyer.
\newblock Progress towards practical quantum variational algorithms.
\newblock {\em Physical Review A}, 92(4), oct 2015.

\bibitem{Wiersema_2020}
Roeland Wiersema, Cunlu Zhou, Yvette de~Sereville, Juan~Felipe Carrasquilla,
  Yong~Baek Kim, and Henry Yuen.
\newblock Exploring entanglement and optimization within the hamiltonian
  variational ansatz.
\newblock {\em {PRX} Quantum}, 1(2), dec 2020.

\bibitem{McClean_2018}
Jarrod~R. McClean, Sergio Boixo, Vadim~N. Smelyanskiy, Ryan Babbush, and
  Hartmut Neven.
\newblock Barren plateaus in quantum neural network training landscapes.
\newblock {\em Nature Communications}, 9(1), nov 2018.

\bibitem{ngsaII}
K.~Deb, A.~Pratap, S.~Agarwal, and T.~Meyarivan.
\newblock A fast and elitist multiobjective genetic algorithm: Nsga-ii.
\newblock {\em IEEE Transactions on Evolutionary Computation}, 6(2):182--197,
  2002.

\bibitem{Deb2011}
Kalyanmoy Deb.
\newblock {\em Multi-objective Optimisation Using Evolutionary Algorithms: An
  Introduction}, pages 3--34.
\newblock Springer London, London, 2011.

\bibitem{nebro2022}
Antonio~J. Nebro, Jesús Galeano-Brajones, Francisco Luna, and Carlos~A.
  Coello~Coello.
\newblock Is nsga-ii ready for large-scale multi-objective optimization?
\newblock {\em Mathematical and Computational Applications}, 27(6), 2022.

\bibitem{Katoch_2021}
S.~Katoch, S.S. Chauhan, and V.~Kumar.
\newblock A review on genetic algorithm: past, present, and future.
\newblock {\em Multimed Tools Appl}, 80:8091–8126, 2021.

\bibitem{pymoo}
J.~{Blank} and K.~{Deb}.
\newblock pymoo: Multi-objective optimization in python.
\newblock {\em IEEE Access}, 8:89497--89509, 2020.

\bibitem{Barkoutsos_2020}
Panagiotis~Kl. Barkoutsos, Giacomo Nannicini, Anton Robert, Ivano Tavernelli,
  and Stefan Woerner.
\newblock Improving variational quantum optimization using {CVaR}.
\newblock {\em Quantum}, 4:256, apr 2020.

\bibitem{Burke2004TheSO}
Edmund~K. Burke, Patrick~De Causmaecker, Greet~Vanden Berghe, and Hendrik~Van
  Landeghem.
\newblock The state of the art of nurse rostering.
\newblock {\em Journal of Scheduling}, 7:441--499, 2004.

\bibitem{Deb1995SimulatedBC}
Kalyanmoy Deb and Ram~Bhushan Agrawal.
\newblock Simulated binary crossover for continuous search space.
\newblock {\em Complex Syst.}, 9, 1995.

\bibitem{Deb1996ACG}
Kalyanmoy Deb and Mayank Goyal.
\newblock A combined genetic adaptive search (geneas) for engineering design.
\newblock {\em Computer Science and Informatics}, 26(4):30--45, 1996.

\bibitem{Spall_1992}
J.C. Spall.
\newblock Multivariate stochastic approximation using a simultaneous
  perturbation gradient approximation.
\newblock {\em IEEE Transactions on Automatic Control}, 37(3):332--341, 1992.

\end{thebibliography}

\onecolumn\newpage
\appendix

\section{Penalty hyperparameter tunning}
\label{append_penaltiestunning}

\begin{figure}[t]
  \centering
\includegraphics[width=\linewidth]{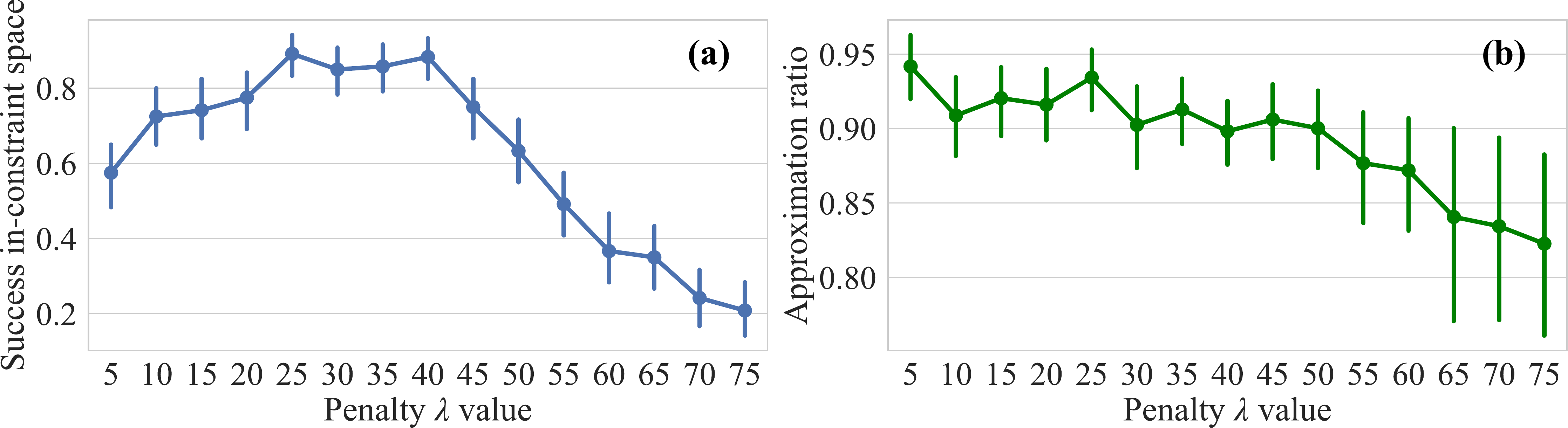}
\caption{Comparison of VQE performance optimizing the CMP penalized cost function~\eqref{eq_penham} with different hyperparameters $\lambda_l=\lambda_f$ values. The results show the average from 120 random instances with two cash points optimized within a four-day period after 500 iterations of the algorithm with SPSA. (a) Percentage of instances whose almost all sampled solutions are within the feasible space ($P(\vec{\theta})>0.99$). (b) Average of the approximation ratios (taking into account only instances with $P(\vec{\theta})>0.99$). Both plots display the $95\%$ confidence interval in the average estimation. (Color online).}
  \label{fig_penalties}
\end{figure}

\begin{figure*}
  \centering
\includegraphics[width=0.935\linewidth]{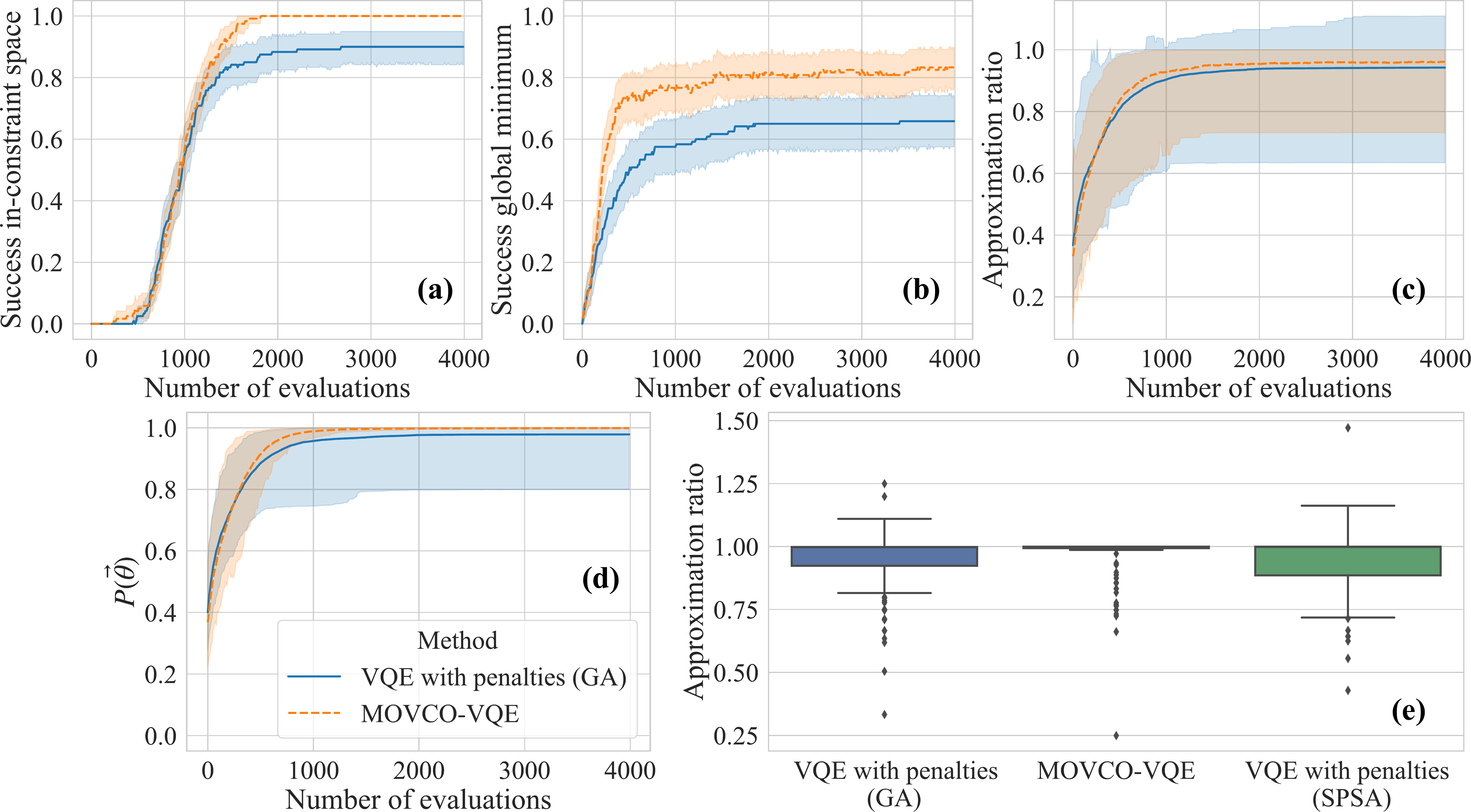}
\caption{Comparison between the MOVCO-VQE and VQE with penalties performance. We show the results from 120 CMP instances (2 cash points within a period of 4 days). In the VQE with penalties, a single objective genetic algorithm with the same genetic operators and hyperparameters as MOVCO is used as classical optimizer. Plots (c) and (d) show a $95\%$ percentile interval around the average ranging from 2.5 to 97.5 percentiles of the distribution. Plots (a) and (b) displays the $95\%$ confidence interval in the average estimation. The boxes in plot (e) show the lower and the upper quartiles, and the whiskers extend to 1.5 interquartile ranges (Color online).}
  \label{fig_MOVQEvsVQE_results_withGA}
\end{figure*}

As explained in the paper, a typical approach to solving combinatorial optimization problems with hard constraints is to include penalty terms in the cost function, as is done in eq.~\eqref{eq_penham}. The penalty hyperparameters $\lambda_l$ and $\lambda_f$ control the gap between feasible and infeasible solutions, and their value affects the performance of the algorithm. A low value of these hyperparameters may speed up the convergence of the algorithm, but it also increases the possibility of getting trapped in a local minimum, and vice versa.  

To increase the significance of the comparison conducted in sec.~\ref{sec_MOVCOVSpenVQE} between MOVCO-VQE and penalized VQE, we numerically studied how the value of these hyperparameters affects the average penalized VQE performance and selected the best ones. We performed 500 iterations of the classically simulated algorithm on 120 randomly generated instances of the Cash Management problem with two cash points within a four days period with different values of the hyperparameters. As in sec.~\ref{sec_MOVCOVSpenVQE}, we used the ansatz~\eqref{eq_ryansatz}, the SPSA optimizer, and 8192 shots. Since $\lambda_l$ and $\lambda_f$ have no qualitative differences we chose $\lambda_l=\lambda_f$. We show the results in Fig.~\ref{fig_penalties}, where we compare the percentage of instances whose almost all sampled solutions are within the feasible space and the approximation ratio obtained. We note that $\lambda_l=\lambda_f=25$ achieved the best results.      

\section{Cash Management instances generation}
\label{append_problemgeneration}

In the computational experiments shown in section~\ref{sec_numresults} the Cash Management instances with $C$ cash points were randomly generated following the next procedure: 
\begin{itemize}
    \item The price of sending or withdrawing cash at the branch or ATM $c$, $k_c$, is an integer drawn from the discrete uniform distribution in the interval $\left[1,4\right]$. The price on the first day is double, $k^0_c = 2k_c$.
    \item The initial prediction of cash available in every cash point $c$ and day $d$, $p_{cd}$, is an integer drawn from the discrete uniform distribution in the interval $\left[-2,5\right]$.
    \item the maximum total cash in the network of branches and ATMs on the last day of the time interval is established in $v_f = C$.
    \item The maximum number of transactions (shipments and withdrawals) that can be
    made each day is $l=1$ if $C=2$, $l= \lfloor \frac{3}{4} C \rceil$ otherwise (where $\lfloor \cdot \rceil$ denotes the rounding function to the nearest integer value).    
\end{itemize}      
As explained in section~\ref{sec_cashmanagement}, the total number of hard constraints is $D+1$, excluding those constraints that are satisfied by any state given the formulation of the problem (lower and upper bound to the available cash in each ATM each day). However, for the small-size problems generated in section~\ref{sec_numresultsMOVCO} and \ref{sec_MOVCOVSpenVQE} it could happen that no state satisfies all the constraints. In such cases, the maximum number of constraints satisfied by at least one solution is calculated, and the restricted energy and constraint satisfaction results are computed based on this value.

\section{Single objective genetic algorithm optimization}
\label{append_singleobjectiveGAoptimization}

After observing the performance of MOVCO, one may wonder whether its advantage is intrinsically due to the constraint multiobjective approach, or whether it is only a consequence of the optimization mechanism of the genetic algorithm. 

In this regard, in Fig.\ref{fig_MOVQEvsVQE_results_withGA} we compare its performance with a VQE in which the classical optimization of the variational parameters is conducted by a single objective genetic algorithm and a cost function with penalty terms. MOVCO-VQE utilizes the NSGA-II algorithm as an optimization subroutine, as explained in sec.~\ref{sec_MOVCO}. In the shown numerical results NSGA-II is implemented with simulated binary crossover and polynomial mutation as genetic operators, and a population size equal to 10. In order to fairly compare the two approaches, in Fig.\ref{fig_MOVQEvsVQE_results_withGA} the VQE with penalties uses a genetic algorithm with the same genetic operators and hyperparameters as the NSGA-II of MOVCO-VQE, but instead of promoting solutions according to the multiobjective optimization of the two fitness functions introduced in sec.~\ref{sec_MOVCO}, the only quality metric is the energy of the penalized Hamiltonian defined in~\eqref{eq_penham}: $\bra{\Psi(\vec\theta)}\hat{C}_{pen}(\hat{Z})\ket{\Psi(\vec\theta)}$. Therefore, the MOVCO ranking based on dominance fronts is replaced by a rating based on the average penalized energy value with $\lambda_l=\lambda_f=25$.  

We can observe how the CMP results obtained using the single objective genetic algorithm are similar to those obtained with the SPSA optimizer (see Fig.~\ref{fig_MOVQEvsVQE_results}), showing an advantage for MOVCO both in the cost of the achieved solutions, in the dispersion of the results, and especially in the ability to converge to solutions that satisfy all the hard constraints. These empirical results indicate that the advantage of MOVCO lies in the combination of the multiobjective approach that simultaneously addresses both the energy of the solutions and the satisfaction of complex constraints in the optimization process, and the effective exploration of the Hilbert space thanks to the use of a genetic algorithm. 

\section{NSGA-II algorithm}
\label{append_NGSAIIpseudocodes}

In this section, we provide the details of the Non-dominated Sorting Genetic Algorithm (NSGA-II)~\cite{ngsaII}, including the general steps of the algorithm (Algorithm~\ref{alg_nsga2}), the classification of the solutions into fronts (Algorithm~\ref{alg_frontclass}), and the calculation of the crowding distance (Algorithm~\ref{alg_crowdingdistance}).

\begin{algorithm}[H]
\caption{NSGA-II Algorithm}\label{alg_nsga2}
 \hspace*{\algorithmicindent} \textbf{Input:}
 population size $N$, number of generations $T$, two fitness functions, genetic operators (selection, crossover, mutation)\\
 \hspace*{\algorithmicindent} \textbf{Output:} a set of solutions that compose the Pareto front
\begin{algorithmic}[1]
\State $P(0)$ $\gets$ RandomPopulation(N) \Comment{Generate the initial population}
\State Evaluate($P(0)$) \Comment{Assign a fitness or rank using non-dominated sorting (see Algorithm~\ref{alg_frontclass}) and then crowding distance operator (see Algorithm~\ref{alg_crowdingdistance})}
\State $Q$ $\gets$ $\emptyset$ \Comment{Initialize offspring population with an empty set}
\For{$t = 0$ to $T-1$}
    \State $W(t)$ $\gets$ TournamentSelection($P(t)$) \Comment{compare individuals two by two and pick the winners}
    \State $M(t)$ $\gets$ CrossoverSBX($W(t)$) \Comment{Allowing to explore the search space creating new solutions~\cite{Deb1995SimulatedBC}}
    \State $Q(t)$ $\gets$ PolinomialMutation($M(t)$) \Comment{Exploitation~\cite{Deb1996ACG}}
    \State Evaluate($Q(t)$) \Comment{Evaluating the new population members}
    \State $P_{new}(t)$ $\gets$ $P(t)$ $\cup$ $Q(t)$ \Comment{Merge population calculating fitness}
    \State $P(t+1)$ $\gets$ Survivor($P_{new}(t)$) \Comment{Choosing better N solutions for next generation}
\EndFor
\end{algorithmic}
\end{algorithm}

\begin{algorithm}[H]
\caption{Crowding Distance Calculation}\label{alg_crowdingdistance}
\hspace*{\algorithmicindent} \textbf{Input:}
 a set of solutions composing the front $F$, objective functions $f_m$\\
 \hspace*{\algorithmicindent} \textbf{Output:} 
 an ordering of the solutions in terms of a density-based distance $d_i$
\begin{algorithmic}[1]
\State $r$ $\gets$ $|F|$ \Comment{Cardinality of $F$ i.e.  number of solutions in front $F$}
\State $M$ $\gets$ number of objective functions 
\For{$i=1$ to $r$}
    \State $d_i$ $\gets$ 0 \Comment{for each element of front $F$ set its distance $d_i$ to zero}
\EndFor
\For{$m=0$ to $M$}
    \State $F$ $\gets$ sort($F$, $f_m$) \Comment{sort the elements $i$ of front $F$ according to their objective function $f_m(i)$ values}
    \State $d_1$ $\gets$ $d_r$ $\gets$ $\infty$ \Comment{set the distance $d$ of the most extreme solutions to large values}
    \State $f_{m,max}$ $\gets$ $\max f_m(i)$
    \State $f_{m,min}$ $\gets$ $\min f_m(i)$
    \For{$i$ = 2 to $r$ - 1}
        \State $d_i$ $\gets$ $d_i$ + $|f_m(i+1) - f_m(i-1)|/(f_{m,max} - f_{m,min})$ \Comment{calculate the distance of each solution as the sum of the differences of the values of the objective functions $f_m$ of the previous and the next element in the sorted front $F$}
    \EndFor
\EndFor \Comment{solutions in less dense spaces, i.e., solutions with higher $d$, are promoted.}
\end{algorithmic}
\end{algorithm}

\begin{algorithm}[H]
\caption{Front classification}\label{alg_frontclass}
\hspace*{\algorithmicindent} \textbf{Input:}
 a set of solutions or population $P$, fitness functions\\
 \hspace*{\algorithmicindent} \textbf{Output:} an ordering of the solutions in terms of Pareto dominance
\begin{algorithmic}[1]
\State $F_1$ $\gets$ $\emptyset$ \Comment{Initialize the first front with an empty set}
\For{each $p$ in $P$}
    \State $S_p$ $\gets$ $\emptyset$, $n_p$ $\gets$ 0
    \For{each $q \neq p$ in $P$}
        \If{$p < q$} \Comment{if $p$ dominates $q$ according to the fitness functions}
            \State $S_p$ $\gets$ $S_p$ $\cup$ $\{q\}$ \Comment{space dominated by $p$}        
        \Else
            \If{$q < p$} \Comment{if $q$ dominates $p$}
            \State $n_p$ $\gets$ $n_p$ + 1 \Comment{increasing the space that dominates $p$}
            \EndIf
        \EndIf
    \EndFor
    \If{$n_p$ = 0}
        \State $p_{rank}$ $\gets$ 1
        \State $F_1$ $\gets$ $F_1$ $\cup$ $\{p\}$ \Comment{construction of the first front}
    \EndIf
\EndFor

\State $i$ $\gets$ 1
\While{$F_i$ $\neq$ $\emptyset$} 
    \State $F_{i+1}$ $\gets$ $\emptyset$ \Comment{construction of the rest of the fronts}
    \For{each $p$ in $F_i$}
        \For{each $q$ in $S_p$} \Comment{each element dominated by $p$}
            \State $n_q$ $\gets$ $n_q$ - 1 \Comment{the space that dominates $q$ is decremented in one element}
            \If{$n_q$ = 0} \Comment{$q$ is now a non-dominated solution, therefore it belongs to the next front}
                \State $q_{rank}$ $\gets$ $i$ + 1
                \State $F_{i+1}$ $\gets$ $F_{i+1}$ $\cup$ $\{q\}$
            \EndIf
        \EndFor
    \EndFor
    \State $i$ $\gets$ $i$ + 1
\EndWhile \Comment{lower ranking solutions, i.e. belonging to the first fronts, are promoted.}
\end{algorithmic}
\end{algorithm}

\end{document}